\newcommand{\homedir}{./hardware_pca_onn}
\makeatletter
\def\input@path{{\homedir/stys_TPAMI/}}
\makeatother

\documentclass[10pt, journal]{IEEEtran}

\ifCLASSOPTIONcompsoc
  \usepackage[nocompress]{cite}
\else
  \usepackage{cite}
\fi

\ifCLASSINFOpdf
  
\else
 
\fi

\newif\ifpeerreview
\peerreviewtrue
\peerreviewfalse

\usepackage{amsmath}
\usepackage{amsfonts}
\usepackage{amssymb}
\usepackage[switch]{lineno}

\newcommand{\papertitle}{
An approach to Physics Based Single Photon Vision System Design
}
\usepackage{enumitem,kantlipsum}

\usepackage{siunitx}

\usepackage[dvipsnames]{xcolor} 
\newcommand{\rnote}[1]{\textcolor{red}{#1}}
\newcommand{\bnote}[1]{\textcolor{blue}{#1}}
\expandafter\let\expandafter\oldequationstar\csname equation*\endcsname%
\expandafter\let\expandafter\endoldequationstar\csname endequation*\endcsname%
\renewenvironment{equation*}%
  {\linenomath\oldequationstar}{\endoldequationstar\endlinenomath}%

\newcommand{\B}{\boldsymbol}
\newcommand{\IR}{\mathbb{R}}
\newcommand{\IZ}{\mathbb{Z}}

\usepackage{subcaption}

\newcommand{\Xalign}[1]{\begin{aligned}#1\end{aligned}}

\newcommand{\Xequa}[1]{\begin{equation}\Xalign{#1}\end{equation}}

\def\Xhide#1{}

\def\noisy#1{\tilde{#1}}
\def\AGN{\mathcal{N}}
\def\Poisson{\mathcal{P}}
\def\Poisson{\mathrm{Poisson}}

\usepackage{graphicx} % include PDF figures
\def\Xpolish#1#2{#2}

\def\bnote#1{\textcolor{blue}{#1}}
\def\rnote#1{\textcolor{red} {#1}}
\def\ynote#1{\textcolor{orange}{#1}}

\def\YLnote{\rnote}

\def\ynote#1{\color{orange} #1 \color{black}}
\def\AVnote{\ynote}

\usepackage[normalem]{ulem}
\def\checkednote#1{\sout{#1}}

\usepackage{tikz}
\usetikzlibrary{shapes.geometric}
\usetikzlibrary {shapes.misc}
\usepackage{caption}
\usepackage{amsmath,amsfonts}

\ifpeerreview
    \linenumbers\linenumbersep 15pt\relax    
    \def\checkednote#1{} % remove comments
    \def\YLnote{}
\else
\fi
\def\YLnote{}
\def\checkednote#1{}

\usepackage{printlen}
\usepackage{layouts}

\begin{document}
\title{\papertitle}

% \author{\IEEEauthorblockN{Michael Shell}
% \IEEEauthorblockA{School of Electrical and\\Computer Engineering\\
% Georgia Institute of Technology\\
% Atlanta, Georgia 30332--0250\\
% Email: http://www.michaelshell.org/contact.html}
% \and
% \IEEEauthorblockN{Homer Simpson}
% \IEEEauthorblockA{Twentieth Century Fox\\
% Email: homer@thesimpsons.com}
% \and
% \IEEEauthorblockN{James Kirk\\ and Montgomery Scott}
% \IEEEauthorblockA{Starfleet Academy\\
% San Francisco, California 96678-2391\\
% Telephone: (800) 555--1212\\

\author{
    \IEEEauthorblockN{Yizhou~Lu\IEEEauthorrefmark{1}, Trevor~Seets\IEEEauthorrefmark{1}, Felipe~Gutierrez-Barragan\IEEEauthorrefmark{2}, Ehsan~Ahmadi\IEEEauthorrefmark{1} and Andreas~Velten\IEEEauthorrefmark{1}\IEEEauthorrefmark{3}\thanks{Andreas Velten is the corresponding author. E-mail:velten@wisc.edu}}\\
    \IEEEauthorblockA{\IEEEauthorrefmark{1}Department of Electrical and Computer Engineering, University of Wisconsin-Madison, Madison, United States\\
    % Email: ylu289@wisc.edu
    }
    \IEEEauthorblockA{\IEEEauthorrefmark{2}Department of Computer Sciences, University of Wisconsin-Madison, Madison, United States\\
    % Email: author2@example.com
    }
    \IEEEauthorblockA{\IEEEauthorrefmark{3}Department of Biostatistics and Medical Informatics, University of Wisconsin-Madison, Madison, United States\\
    % Email: author2@example.com
    }
    
}

% \author{

% \Bingnote{AUTHORS}

% }


% compsoc conferences compared to traditional conferences), use this
%Telephone: (800) 555--1212, Fax: (888) 555--1212}



\maketitle
% title is in include.tex

\pagestyle{plain}
\pagenumbering{arabic}
\begin{abstract}
% The abstract goes here.
The design of the camera and optical measurement is a crucial part of optimizing machine vision systems.
However, camera designs are usually optimized to produce human-interpretable images.
Moreover, camera optimization typically makes the assumption of additive noise, while modern optical imaging systems are mainly affected by photon noise which is not additive. 
Previous studies have highlighted the fundamental effect of the noise model on the outcome of the design process in the context of coded single-pixel and compressed sensing cameras. 

In addition to the exact noise model, the nature of the data encountered and the vision task fundamentally affect the optimal hardware design. 
This means that a camera or compressed sensing code that is optimal for capturing images is almost certainly not optimal for any downstream vision task.
In this work, we used a simple end-to-end model on a single-pixel camera to study the effects of the noise model for different vision tasks.
We demonstrate the importance of incorporating non-linear noise features in optimizing even the simplest vision systems. 

\end{abstract}



\ifCLASSOPTIONpeerreview
\begin{center} \bfseries EDICS Category: 3-BBND \end{center}
\fi
\IEEEpeerreviewmaketitle

% \input{include}

\iffalse
% this is the older version
\noindent \textbf{\bnote{Notes for Paper Contributors:}}

\begin{itemize}
    \item To define new commands or include additional packages, add to \textit{include.tex}. If we have to change latex template this minimizes effort.
    \item To add new sections, add them under \textit{sections} and then edit \textit{main.tex}.
    \item For examples of how to add Figures, Tables, see \textit{examplelatex\_iccp2020\_template}.
    \item One method to create bibtex entries is by going to google scholar, searching paper title, clicking on the cite icon, clicking on Bibtex, and copying and pasting the google scholar bibtex to \textit{references.bib}
\end{itemize}

\rnote{
% 1. Make noiseless classification result 
2. Select data for ONN experiment 
3. Replace Random Hadamard by Compressed Sensing ?
3. Split decoder and encoder in ONN
4. Show regularization improves all masks (equally), then I can use random truncated 5. Supplementary includes how the experiment was done 6. The name is 
\textbf{Selective Sensing} 7. Use Seaborn theme 
}

\medskip
\medskip
\medskip
\medskip
\medskip
\medskip

%%%%%%%%% BODY TEXT

% The first section title should be wrapped inside a \IEEEraisesectionheading as follows.

% \IEEEraisesectionheading{
%   \section{Introduction}\label{sec:1_intro}
% }

% \printlength\linewidth
\printinunitsof{in}\prntlen{\linewidth}

\noindent\rule{\linewidth}{0.4pt}
\fi

    \YLnote{\checkednote{1. Replace Low frequency Hadamard with random frequency Hadamard 2. Fulfill the paragraphs left}}

\section{Introduction}\label{sec:1_intro}
\textbf{\YLnote{
\checkednote{Feedbacks:  
1. The methodology is straightforward and seems incomplete.
The key question this paper hopes to answer is the optimized coding strategy under Poisson noise, but it uses a Gaussian noise assumption (though with reparameterization) to train the proposed model. The downstream tasks (classifying only 10 numbers) are oversimplified as well. All the future improvements as mentioned in the limitations are quite important factors to make this work complete.
2. The experiments are insufficient.
There is no comparison with any relevant method proposed in recent years. All coding strategies analyzed in Fig. 7 are classic ones.
3. The relevance to TPAMI is also poor. There is no citation to any relevant papers in earlier or recent issues from TPAMI. As a manuscript submitted to the regular track, the length is also a bit too short (only 9 pages excluding reference and bios).}}}

% outline
\iffalse

1. why we need co-design of hardware and software, and and use single pixel camera to demonstrate

2. what is a single pixel camera, what is coding, decoding; a bad coding can degrade the performance

3. a key factor in finding optimal coding is noise/ if no noise, no coding is needed

4. coding for poisson noise is different from coding for AGN

5. -

\fi

\iffalse
\IEEEPARstart{T}{his} demo file is intended to serve as a ``starter
file'' for ICCP 2020 submissions produced under
\LaTeX~\cite{kopka-latex} using IEEEtran.cls version 1.8b and later.
\fi

\IEEEPARstart{W}{}hile we can mass-produce optical components of increasing complexity, most vision hardware still closely mimics the optics of the human eye to create easily interpretable images. 
In this approach, the lens projects an image that is then post-processed to extract patterns.
However, more generalized vision systems can perform alternative measurements that may be more perceptually meaningful than simply producing images. 
Designing such systems requires a collaborative effort to integrate hardware and digital processing algorithms while considering the appropriate noise model.
The significance of this approach is particularly illustrated by the single-pixel camera in this context.

The single pixel camera uses the concept of optical coding, or multiplexing, which involves projecting an image onto a mask and collecting the transmitted light with a large sensor pixel.
The system can potentially achieve an improved signal-to-noise ratio (SNR) and higher light throughput by coding compared to point-by-point measurements \cite{mitra2014can}.
In this context, a measurement vector $\B{y}$ is formed from a set of measured photon counts, with the corresponding masks vectorized as rows of the sensing matrix $\B{M}$.
Ignoring the noise, coding can be represented as
 \begin{equation}
    \label{eqn:NoiselessMeasurement}
    \begin{aligned}
        \B{y} &=  \B{M x} ,
    \end{aligned}
\end{equation}
where $\B{x}$ is the signal vector. 
Decoding refers to the process of recovering the initial signal $\B{x}$ from $\B{y}$.
The recovery quality of $\B{x}$
is significantly dependent on the conditioning of $\B{M}$ and the light throughput of $\B{M}$~\cite{mitra2014can}.
Suboptimal $\B{M}$ have the potential to significantly degrade the performance of the vision system during the recovery process, as highlighted by Mitra et al. \cite{mitra2014can}. 
%The optimization of coding schemes regarding $\B{M}$ is intricately linked to the characteristics of noise. 

While a greater light throughput, or a lower noise level, can improve recovery, the improper choice of the noise model can also give rise to a suboptimal coding at specific noise levels. 
Coding design is often performed under the assumption of signal-independent additive Gaussian Noise (AGN) which arises from imperfect sensors.
With advances in sensor design, even the sensor noise of low-end camera sensors has been reduced to a few photons per pixel.
This reduction is significant enough so that nonadditive photon noise becomes the primary noise factor in the vast majority of imaging applications \cite{cossairt2012does}. Photon noise follows a super-Poissonian distribution~\cite{mandel1959fluctuations}. For passive cameras operated with incoherent ambient light the noise can usually be approximated as Poissonian.

The transition from AGN to more realistic photon noise fundamentally alters the designs of vision systems.
For example, although single-pixel cameras with random masks provide performance advantages under AGN, they do not outperform a simple sequential scan when dealing with Poisson noise \cite{harwit1979hadamard, swift1976hadamard, willet2009CSPoisson, scotte2022photon_noise, vanden2019various}.
Moreover, optimal coding designs under AGN conditions do not deliver optimal performance under Poisson\Xhide{photon} noise at the same noise levels.

To address this issue, Mitra et al. \cite{mitra2014can} develop a method using data-driven priors to design optimized codes for image reconstructions. 
Their approach significantly improves the performance of coding under Poisson noise by designing measurements specifically targeting sparsity or compressibility in the data \cite{mitra2014can}. 
% It can markedly enhance the reconstruction performance of coding schemes under Poisson noise \cite{mitra2014can} by designing measurements specifically targeting sparsity or compressibility in the data. 
However, image recovery is typically not the ultimate goal for modern vision systems.
Since vision tasks, such as classification, operate in embeddings created from images, their data should be inherently more compressible than their upstream imaging tasks, allowing for more effective measurement designs. 
To show this, we implement a simple end-to-end vision system that can be optimized under different noise assumptions. 
We illustrate why AGN is not suitable for optimizing coding under \Xpolish{Poisson}{photon} noise and show that a coding scheme optimized for a specific vision task and for the correct noise model can perform close to the optimal non-coding camera even when \Xhide{only }photon noise is present.

Our contributions are as listed.
\begin{itemize} 
\item {We introduce the methodology of Selective Sensing (SS), which encompasses coding techniques specifically designed to extract data-driven priors selectively for \YLnote{various downstream tasks.} \Xhide{\YLnote{These coding methods can be seamlessly integrated with models for downstream tasks, providing a novel approach to efficiently incorporate learned priors into various applications.\checkednote{\textbf{sales pitch?}}}}}
\item {
We provide an end-to-end vision model using classification as the performance metric, which is more perceptually meaningful than mean-squared-error (MSE), to optimize $\B{M}$.
}
\item {
We propose a hybrid optimization method for non-differentiable Poisson noise that employs AGN with signal-dependent variance and the reparameterization trick to estimate gradients during training, while switching to Poisson noise in the testing stage for model assessment.
}
\end{itemize}

Our model is extendable to more general camera optics and different vision tasks in future work.

% \hrule

% \input{\homedir/sections/1_intro/1_intro_2024_09_29.tex}

% \input{\homedir/sections/1_intro/1_intro_2024_04_11.tex}

% \input{\homedir/sections/1_intro/1_intro_2024_04_01.tex}

% \input{\homedir/sections/1_intro/1_intro_2024_03_02.tex}

% \input{\homedir/sections/1_intro/1_intro_2024_01_03.tex}

\section{Related Work}
\label{sec:2_related}

\textbf{Coding under \Xpolish{Poisson}{photon} noise}. In coded imaging, \Xhide{coding allows the capture of a two-dimensional image with a single-pixel sensor \cite{harwit1979hadamard}. }Hadamard matrices are considered the optimal coding scheme for multiplexing \cite{harwit1979hadamard, cossairt2012does, wuttig2005optimal, mitra2014can} in systems with only AGN, although this conclusion does not hold for photon noise \cite{mitra2014can}. 
\Xpolish{However, when Poisson noise is predominant, we should not use any coding \cite{harwit1979hadamard, cossairt2012does}. When the noise is Poisson distributed, coding with sparse priors such as compressed sensing is not recommended \cite{willet2009CSPoisson, willett2011poisson, vanden2019various}.}
{In many recent projects, optimization efforts focus on algorithmic enhancements for reconstruction, often maintaining the use of random coding strategies \cite{goyal2016performance, gibson2020single, liutkus2014imaging, candes2008introduction}\YLnote{\checkednote{need more here}}, though this type of matrices are not recommended under \Xpolish{Poisson}{photon} noise\cite{willet2009CSPoisson,  willett2011poisson, vanden2019various}.
% \Xhide{ While some approaches, such as Feature Specific Imaging (FSI), optimize sensing matrices based on learnable priors \cite{neifeld2003dual, neifeld2003FSI, neifeld2014optimizing}, these methods explicitly do not consider Poisson noise. Thus, the challenge of optimizing masks under Poisson noise persists \cite{neifeld2014optimizing}.}
\YLnote{Wuttig and Ratner, et al. \cite{wuttig2005optimal, ratner2007optimal} introduced \Xpolish{Poisson}{photon}-noise-aware optimization techniques for \Xpolish{multiplexing}{coding} matrices that outperform Hadamard in the presence of moderate signal-dependent noise. 
However, their analysis neglects the data priors, leading to suboptimal matrices for greater \Xpolish{Poisson}{photon} noise \cite{mitra2014can}.} Additionally, attempts at matrix optimization focusing on minimal mutual coherence \cite{mordechay2014matrixRIP} \Xpolish{lack a foundation in Poisson noise model}{do not adequately consider photon noise}, as highlighted in this study.}

% \textbf{Task Specific Imaging}

\textbf{Feature Specific Imaging}. Feature-specific imaging (FSI) is a type of imaging system \Xpolish{that directly measures linear features of the object irradiance distribution, instead of forming a conventional image and then extracting features from it}{where sensing matrix $\B{M}$ incorporates linear features of expected data} \cite{neifeld2003FSI, neifeld2003dual}. This approach can provide higher feature fidelity and lower detector count than conventional imaging, especially for applications that require relatively few features \cite{neifeld2003FSI, neifeld2003dual, neifeld2014optimizing}. This technique can be viewed as a variant of compressed sensing, in which the sensing matrix is determined based on prior information \cite{neifeld2014optimizing}. However, \Xpolish{its performance and implications under Poisson noise conditions remain relatively unexplored in the existing literature, warranting further discussion and investigation}{\YLnote{current state-of-the-art} FSI is explicitly designed for AGN rather than Poisson noise\cite{neifeld2014optimizing}}. \YLnote{FSI can be regarded as a technique that leverages data driven priors, and Mitra et al. have asserted that employing data driven priors can enhance image recovery under Poisson noise \cite{mitra2014can}. Therefore, in our work, we underscore the promising prospects of extending feature-specific concepts to a broad spectrum of vision tasks under Poisson noise.}

\textbf{End-to-end optimization}. This method \Xpolish{refers to training hardware and software networks for image processing pipelines}{jointly optimizes camera design and digital signal processing to achieve optimal overall performance} \cite{diamond2021dirty, zhang2021deep, jacome2023middle}. \YLnote{In many previous projects, this idea was usually implemented \Xpolish{without considering Poisson noise}{without considering Poisson noise in cost functions} \cite{hinojosa2021learning, dun2020learned, metzler2020deep, chang2019deep, onzon2021neural, spall22hybrid_training, jacome2023middle, gibson2020single} or without optimizing masks \Xpolish{under Poisson noise during training}{with gradient incorporating Poisson distribution} \cite{tseng2021differentiable, diamond2021dirty, rego2022deep, duarte2008CS, nature2022ONN, gibson2020single, xu2020compressed}. Rego et. al froze the sensing matrix as a pinhole setting without optimizing it \cite{rego2022deep}}.
Wang et al. \cite{nature2022ONN} successfully implemented a neural network model for handwritten number classification on an optical device with limited photon budget, demonstrating the potential for AI-assisted optimization of coding schemes\Xhide{ in CI}. 
\YLnote{However, Poisson noise was considered only in model testing where the most robust model was picked among a set of \Xpolish{hyper-parameter combinations}{different initialization seeds} \cite{nature2022ONN}. %supplementary note 13
Xu et al. \cite{xu2020compressed} examined the robustness of the proposed neural network model to Poisson noise; however, the training process does not explicitly incorporate the gradients affected by Poisson noise.}
As we see later, however, the correct noise model is essential to obtain an optimal or near-optimal optical design.
Our contribution in this area is developing a \Xpolish{noise-included}{noise-aware} training approach within a neural network model to find globally optimal masks under Poisson noise. 
\YLnote{\Xpolish{
% these part is rewritten to avoid explaining the details
Addressing the limitations identified by Mitra et al. \cite{mitra2014can}, this model presents the following advantages.
    \begin{itemize}
        \item The optimization of coding finds a globally optimal $\B{M}$ rather than its greedy approximation.
        \item Noise is incorporated during the training process to simulate real-world conditions instead of using a one-dimensional affine noise model.
        \item The training process is accelerated and computationally more efficient as the binary constraint is addressed even in real experiments. Additionally, the prior is learned from the entirety of the light fields and spectral data, rather than from patches.
        \item The contribution of a data-driven prior to performance enhancement is more accurately discerned using the classification rate as the metric.
    \end{itemize}
}
{
Although our approach also employs data-driven priors, which is similar to that of Mitra et al., our research extends beyond simple image reconstruction to address a wider spectrum of vision tasks and investigates task-specific optimization techniques on optical hardware.
% Though using data-driven priors as Mitra et.al doing, our research expands its  to encompass a broader range of vision tasks and explores optimization techniques specifically tailored for optical hardware instead of simple image reconstruction.
% Different Mitra's work which primarily uses data-driven priors for image reconstruction, our research expands its  to encompass a broader range of vision tasks and explores optimization techniques specifically tailored for optical hardware.
}
}

% \input{\homedir/sections/2_related/2_related_revised_2}

% \section{Single-Pixel Image Formation Model}
\section{Background}
\label{sec:3_background}
\AVnote{\checkednote{this paragraph has no new information. probaply can be removed}}
\Xhide{\Xpolish{In this work we study the coding problem by considering the single pixel camera as a specific implementation. This system provides a well studied example of coding and is therefore a good toy model to work with. Our findings, however are general and should apply to any form of coding. In this section, we describe some implementation details in our work\ynote{can mention spectral simulations if they are included}}
{In this study, we investigate the coding problem with a specific focus on the single-pixel camera as an illustrative implementation. This system serves as a well-established paradigm for coding, making it an insightful model for our research. Although our exploration is rooted in this particular example, our findings are intended to be of broader applicability, extending to diverse coding forms. In this section, we elaborate on implementation details central to our study.}}

% \Xhide{Our study on coding focuses on the single pixel camera, which serves as a specific implementation that provides a well-studied example of coding. We use this system as a toy model to investigate the problem, but our findings have broader implications and can be generalized to any form of coding.}

\Xpolish{A single-pixel sensor, such as a photon multiplier tube (PMT), is capable of detecting the total number of photons in a measurement but lacks the ability to provide spatial information. In order to obtain this information, a Digital Micromirror Device (DMD) can be used to spatially code the signal. The DMD consists of millions of small tunable mirrors placed at the image plane of the camera objective lens. Each mirror reflects photons photons from a pixel in the image one of two separate detectors. During a measurement, the sensor counts the photons from the DMD
micromirrors, or pixels, that are facing it, providing spatial information about the distribution of photons within the field of view. }
{\Xhide{While a single-pixel sensor like a photon multiplier tube (PMT) can detect the total number of photons in a measurement, it cannot provide spatial information. A single pixel camera uses a Digital Micromirror Device (DMD) located at the image plane of the camera objective lens to code the signal.}}

\subsection{Single Pixel Camera}\label{ssec:singlepixelcamera}

\begin{figure}[ht]
    \centering
    \begin{subfigure}{0.75\linewidth}
        \centering
         \caption{Direct Measurement Strategy: Raster Scan}
        % \includesvg[width=0.637\linewidth]{\homedir/figures/Background/RasterScan2.svg}
        \includegraphics[width=1\linewidth]
        {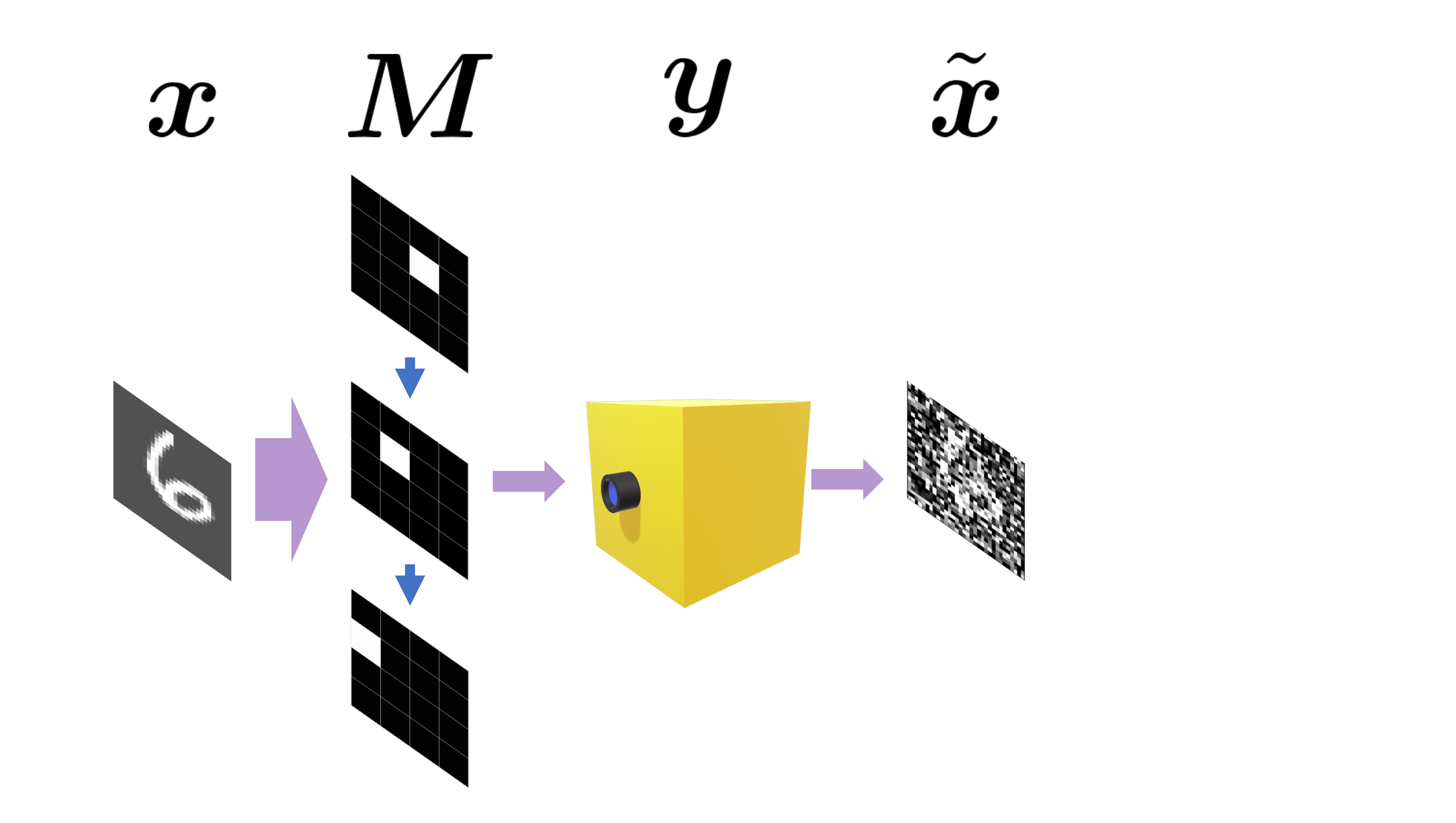}
        % \caption{Non-computational Imaging scheme}
        \label{fig:RasterScan}       
    \end{subfigure}
    \begin{subfigure}{0.75\linewidth}
        \centering     
        \caption{Computational Imaging (CI) Strategy: Basis Scan}
        \includegraphics[width=1\linewidth]{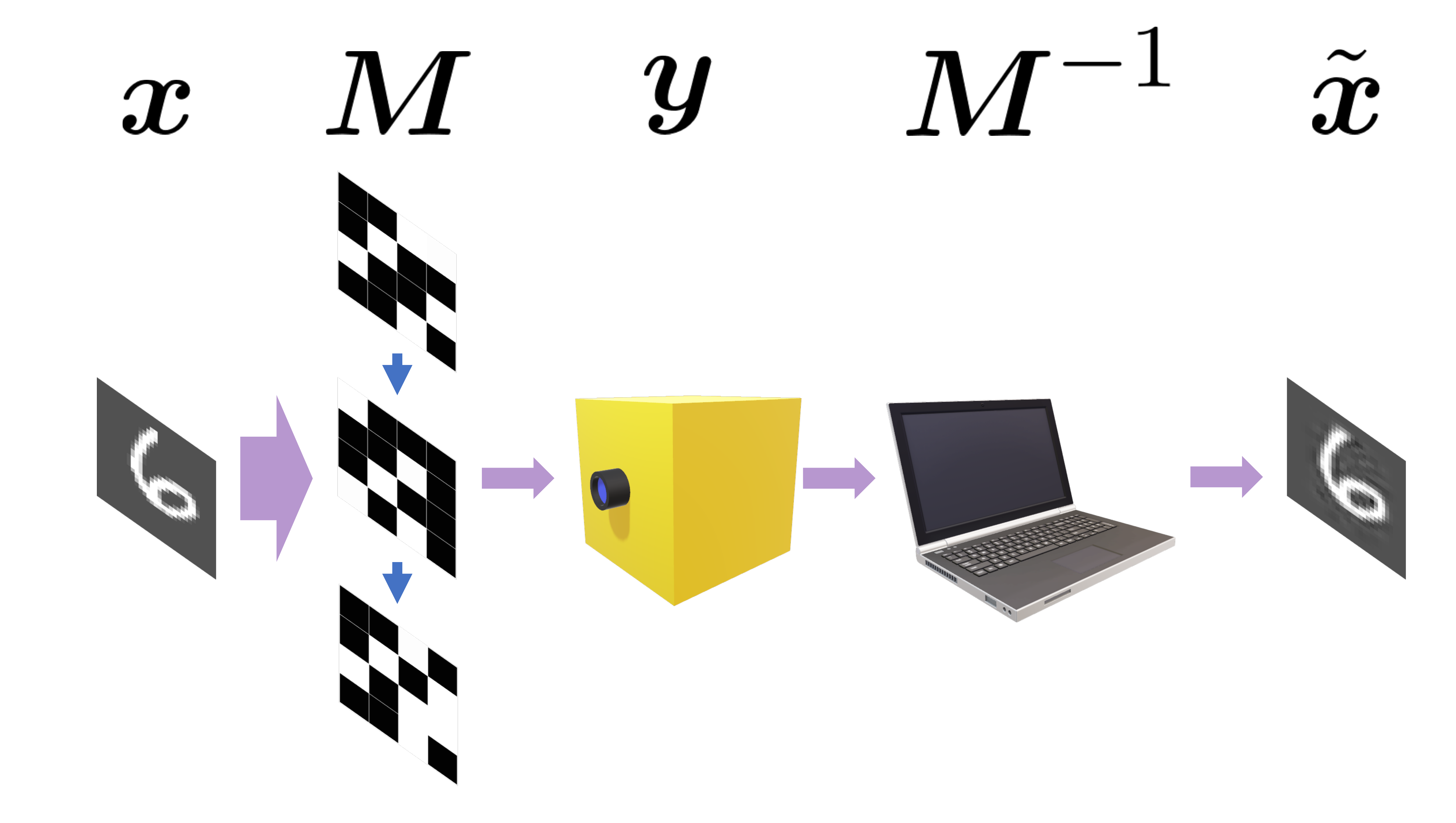}
        % \includesvg[width=0.9\linewidth]{\homedir/figures/Background/BasisScan2.svg}        
        \label{fig:BasisScan}        
    \end{subfigure}
    \caption{$\B{x}$: object, $\B{M}$: masks of coding, $\B{y}$: counted photon numbers at the sensor, $\B{M}^{-1}$: reconstruction operator, $\tilde{\B{x}}$: reconstructed object. (\subref{fig:RasterScan}) Only one pixel (white) is scanned in each measurement, and the measured data requires no reconstruction. (\subref{fig:BasisScan}) The sum of all white pixels is measured in each measurement and it requires a decoding step to reconstruct the field of view.
    }
    \label{fig:SinglePixelImaging}
\end{figure}

The {single-pixel camera} \Xpolish{demonstrated}{shown} in Fig. \ref{fig:SinglePixelImaging} is a \Xpolish{typical}{popular} example of \Xpolish{this model}{coding}, and serves as the primary example in this project. \YLnote{It employs a rapid single-pixel detector in conjunction with a Digital Micromirror Device (DMD) to sequentially multiplex a sensing matrix by modulating various patterns on the DMD \cite{mitra2014can}.} \Xpolish{Coding with a single-pixel-camera mathematically follows the Eq. \ref{eqn:NoiselessMeasurement} while the variables have more concrete meanings as $\B{x} \in \IR^{N \times 1}$ is the image representation of the field of view (FOV) consisting of $N \in \IZ^+$ pixels and is measured by $m \in \IZ^+$ \Xpolish{projections}{measurements, whose photon counts are denoted as $\B{y} \in \IR^{m \times 1}$}. $\B{M} \in \IR^{m \times N}$ is a set of sensing masks linearly projecting the \Xpolish{FOV}{$\B{x}$} onto the sensor pixel\Xhide{, and $\B{y} \in \IR^{m \times 1}$ are the corresponding measured {flux} levels or photon counts }\cite{willet2009CSPoisson}.}{
Its measurement process mathematically follows the Eq. \ref{eqn:NoiselessMeasurement}. Here, $\B{x} \in \IR^{N \times 1}$ is the image representation of the field of view (FOV), which consists of $N$ pixels, and is measured by linear projections or masks, $\B{M} \in \IR^{m \times N}$ to attain corresponding {flux} levels or photon counts, $\B{y} \in \IR^{m \times 1}$ 
 \cite{willet2009CSPoisson}. 
}
% Supposing the image representation of the field of view (FOV) $\B{x} \in \IR^{N \times 1}$ consists of $N \in \IZ^+$ pixels and is measured by $m \in \IZ^+$ \Xpolish{projections}{measurements, whose photon counts are denoted as $\B{y} \in \IR^{m \times 1}$}, 
%   the single-pixel-imaging process, if ignoring noise, can be expressed as the following equation 
%  \begin{equation}
%     \label{eqn:NoiselessMeasurement}
%     \begin{aligned}
%         \B{y} &=  \B{M x} ,
%     \end{aligned}
% \end{equation}
% where \Xhide{$\B{x} \in \IR^{N \times 1}$ is the image representation of the field of view, }$\B{M} \in \IR^{m \times N}$ is a set of sensing masks linearly projecting the \Xpolish{FOV}{$\B{x}$} onto the sensor pixel\Xhide{, and $\B{y} \in \IR^{m \times 1}$ are the corresponding measured {flux} levels or photon counts }\cite{willet2009CSPoisson}. 
Physically, $\B{M}$ can be implemented on the DMD by directing or blocking different parts of the incoming light from $\B{x}$ \cite{raskar2009computational} and averaging them on sensors that digitize the detected flux levels or photon counts $\B{y}$. If the sensing matrix $\B{M}$ is full-rank, \Xpolish{this model accurately characterizes conventional single-pixel imaging scenarios}{the \Xpolish{FOV}{$\B{x}$} can be reconstructed by $\B{M}^{-1}\B{y}$}. 
However, even when $m < N$, \Xpolish{the model remains effective through the integration of regularization techniques, specifically }{we can still recover the \Xpolish{FOV}{$\B{x}$} via} compressed sensing. \YLnote{While Hadamard matrices are considered effective for $\B{M}$, their suitability diminishes in the presence of data-dependent noise \cite{mitra2014can, harwit1979hadamard}.} In this work, we will study the performance of this camera for different choices of $\B{M}$ under different noise models. 
\Xhide{While a compressed sensing system that reconstructs an image fits this paradigm, we will focus on highly compressible problems such as character recognition. In these problems the differences between the common linear Gaussian noise model of the camera and the real Poisson noise model are most apparent. \AVnote{I don't know if this is still needed. We do a good job explaining the goals in the intro Apr 10, 2024 12:03 PM}}

\subsection{Selection of Sensing Matrix}\label{ssec:selectionofsensingmatrix}

% \YLnote{1. What are sensing matrix 2. their types 3. limitations}

% \YLnote{1. What are sensing matrix 2. their types 3. limitations}
\Xhide{As a representative technique in the CI, the mathematical underpinning of single-pixel camera is encapsulated by Eq. \ref{eqn:NoiselessMeasurement}. The diverse selections of the sensing matrix $\B{M}$ lead to distinct coding strategies, as elaborated upon subsequently. }

\YLnote{\Xpolish{\AVnote{ChatGPT problem}In the realm of the Computational Imaging (CI), the foundational equation for single pixel imaging is encapsulated in Eq. \ref{eqn:NoiselessMeasurement}}{Eq. \ref{eqn:NoiselessMeasurement} is the foundational equation for a single-pixel camera}}. 
\Xpolish{The diverse array of choices for the sensing matrix $\B{M}$ engenders two basic coding strategies, as expounded in Fig \ref{fig:SinglePixelImaging}. However, the process of discerning an optimal $\B{M}$, whether drawn from random matrices or Hadamard matrices, relies on multifaceted considerations. In the present study, we advocate the notion that the selection of an ideal $\B{M}$ should be contingent upon the \Xpolish{requisite prior information pertinent}{prior information requisite relevant}  to the tasks at hand. Herein, we delineate various coding types along with their corresponding optimal applications for reference and guidance.}{
Based on it, we distinguish different types of $\B{M}$ below for reference and guidance.
}

\subsubsection{Coding Types}

\begin{itemize}
    \item \textbf{Raster scan} entails a pixel-wise scanning approach where photon counts for all pixels are sequentially measured \cite{gibson2020single}. Mathematically, the sensing matrix $\B{M} = \B{I}$, an identity matrix.      
    
    \item \textbf{Impulse imaging} (conventional imaging) entails an ideal camera that simultaneously captures \Xpolish{complete hyper-spectral data}{all elements in the measurements, $\B{y}$}. 
    % A prominent instance is a pixel array.
    % The most important feature is, It employs a pixel array sensor rather than a single-pixel sensor. 
    % It detects all photons but raster scan only collects $\frac{1}{N}$ photons. Equivalently, each pixel receives $N$ times the exposure in comparison to the raster scan scenario.
    The key feature of this system is its use of a pixel array sensor instead of a single-pixel sensor. This allows for the detection of all incoming photons, whereas the raster scan only collects a small fraction, specifically $\frac{1}{N}$ when the field of view $\B{x}$ consists of $N$ pixels. 
    Therefore, the sensing matrix assumes the form $\B{M} = N\B{I}$ where $\B{I}$ is an identity matrix.      
    This technique sets a benchmark akin to a gold standard, serving as an upper limit for the CI performance and an indicator of the noise level for comparison.
    
    \Xhide{\item \textbf{Non-Selective Codes} \Xpolish{are codes that are not adjusted to the specific sparsity of the dataset. Any general code not generated using the statistics of a representative training set is included in this group. This includes many codes used in practice today, such as Hadmard patterns and Rademacher Random codes. We also include truncated Hadamard patterns that are popular in compressed sensing. A truncated Hadamard code uses a random subset of the full Hadamard basis.
    }{encompass coding schemes that lack adaptation to the inherent sparsity of a given dataset. Codes falling within this category are not crafted based on the statistical attributes of a representative training set. This classification encompasses a range of widely used codes, such as Hadamard patterns and Rademacher Random codes\Xhide{, which do not leverage the specific characteristics of the dataset}. This grouping also extends to truncated Hadamard patterns\Xpolish{, a popular choice in compressed sensing, which employ random subsets of complete Hadamard basis.}{ which employ low frequency components of complete Hadamard basis.}
    }
    }%change to a newer and simpler one
    \item \YLnote{\textbf{Non-Selective Codes} refer to coding schemes that uncorrelated with the inherent \Xpolish{sparsity}{compressibility} or statistical attributes of a specific dataset. Examples include widely used codes like Hadamard patterns and Rademacher Random codes. This category also includes truncated Hadamard patterns that use low-frequency components of the complete Hadamard basis.}

    \item \textbf{Selective Codes}
    \Xpolish{In this work we show that for single photon cameras, codes need to be specifically selected to measure sparse representations of the data. To achieve this we employ PCA codes and ONN generated codes. The principal component analysis (PCA) is a classical method for multispectral data correspondence analysis and classification \cite{rodarmel2002PCA, carr1999PCA, jensen1996PCA, gonzales1987PCA, schowengerdt2006PCA}. In this project, the most significant components are computed from the training data and implemented as the hardware masks. Measured photon counts $\B{y}$ are then be feed to a classifier for model optimization. The optical neural networks (ONN) method is like a fully-connected neural networks model but its first layer is used to simulate the sensing matrix $\B{M}$. By properly applying the physical constraints on the first layer, it is possible for this model to find optimal sensing matrix for a given task.}
    {\Xpolish{.In this study, we elucidate the necessity for the tailored selection of codes in}{encompass coding schemes selectively \YLnote{\Xpolish{incorporating the context of single pixel cameras}{learn the significant features from training data}}} to effectively capture sparse data representations. \Xpolish{To address this, we harness both PCA codes and ONN-generated codes. Principal Component Analysis (PCA) serves as a classical technique for analyzing and classifying multispectral data correspondences \cite{rodarmel2002PCA, carr1999PCA, jensen1996PCA, gonzales1987PCA, schowengerdt2006PCA}. Specifically, we compute the most significant components from training data and implement them as hardware masks within this project. The recorded photon counts, $\B{y}$, are subsequently fed into a classifier to optimize the model. The Optical Neural Networks (ONN) approach mirrors a fully-connected neural network model, yet with its first hidden layer simulating the sensing matrix $\B{M}$. By adeptly applying physical constraints to this foundational layer, the model becomes capable of determining the optimal sensing matrix for a designated task.}{When under AGN, FSI finds data specific codes through Principal Component Analysis (PCA).}
    }
    
\end{itemize}

\subsubsection{Vision Tasks}

\begin{itemize}
    \item Signal Reconstruction (Null-prior/No Compression):
    In tasks involving classic imaging, a null-prior is assumed, and there is no compression during measurements. The rank of the measurement matrix $\B{M}$ is required to be equal to $N$, the number of pixels in the \Xpolish{FOV}{$\B{x}$}\Xpolish{. This scenario, such as in raster scan applications, is universal and does not necessitate prior knowledge of the FOV. Full-rank Hadamard matrices can be employed for reconstruction in noise-free situations. Common coding types for this task include raster scan and non-selective codes.}{ and no prior or bias should be introduced through $\B{M}$.}
    \Xpolish{
    \item Image Reconstruction and Computational Imaging (Weak-prior/Some Compression):
    In this kind of tasks, a weak-prior assumption is made\Xhide{, allowing for some compression during measurements}. This idea is based on prior knowledge, such as signals being sparse in some linear space, with noise concentrated in the high-frequency domain. \Xpolish{An example is the JPEG format. If only the low-frequency components of Hadamard matrices are measured for reconstructing the FOV, this task is engaged. Non-selective codes are typically used for this purpose.}{One example in this category is compressed sensing.}    
    }%hide and update to newer one below    
    {\item Computational Imaging (Weak-prior/Some Compression): In this kind of task, a weak-prior assumption is made. This idea, which focuses on image reconstruction, is based on prior knowledge, such as signals being sparse in some linear space, with noise concentrated in the high-frequency domain. One example in this category is compressed sensing.
    }
    \item Vision (Strong-prior/Much Compression):
    {Many computer vision tasks} involve a strong prior, enabling substantial compression during measurements. 
    \Xpolish{Directly capture of principal features}{Activation of the main feature directly} by $\B{M}$ is possible if some of its components resemble $\B{x}$. In these tasks, a few measurements are often sufficient with a strong prior derived from training data. 
    Our focus is on optimizing the codes in this scenario.

\end{itemize}

\Xhide{\bnote{Use direct measurement / trivial measurement}}

\Xhide{
\begin{enumerate}
    \item Static strategies: 
    Masks are predetermined and fixed. \ynote{aren't they fixed for compressive strategies as well??? Maybe you are trying to distinguish between selective and non-selective masks?}
    \begin{enumerate}
        \item Non-compressible strategies \bnote{(Full rank)}
        \begin{enumerate}
            \item Raster Scan (RS). In this strategy, only one pixel will be scanned in each measurement. Thus, $\B{M} = \B{I}$.
            \item Impulse Imaging (II). II captures the field of view by a pixel array. In this case, each pixel gets $N$ times as much as the exposure time of RS. Mathematically, $\B{M} = N\B{I}$. Alternatively, we can treat it as another RS with $N$ times as many photons. The II works as the baseline for the comparisons with other strategies. 
            \item Hadamard Basis (HB). In this strategy, we choose the 2-D Walsh-Hadamard matrix $\B{H}$ as the mask basis. Hadamard matrices composed of $\pm 1$ has been proved as the best coding strategy \bnote{WD is only mentioned once way earlier, probably better to just change this language.} for reconstruction by Hotelling in 1944 under the AGN.
            \Xhide{\bnote{next part of this sentence is confusing}
            when noise is independent from this signal 
            such as the AGN\cite{cossairt2012does,harwit1979hadamard,wuttig2005optimal}.}
        \end{enumerate}        
        \item Compressible strategies
        \begin{enumerate}
            \item Low-Frequency Truncated Hadamard (LFTH). Given a default Walsh-Hadamard matrix $\B{H}$ and  pick first several rows of it as $\B{M}$. \Xhide{\bnote{switching to low freq}}
            % \item Low-Frequency-Truncated Hadamard (LFTH). Given a default Walsh-Hadamard matrix $\B{H}$ and only keep the first several rows as $\B{M}$.
            % \item Sorted-Truncated Hadamard (STH). Re-order the rows according to their correlation coefficients with the mean of training images. More relevant ones are picked first.
            \item Hardware Principal Components Analysis (HPCA). The DMD displays the PCA components calculated from the training images. \bnote{this learns from data so should maybe go with the ONN? also we should mention this is a novel technique, and give some more explanation to what PCA is}
        \end{enumerate}
    \end{enumerate}    
    \item Selective Sensing\Xhide{Dynamic strategies \bnote{Dynamic is probably not a good name, "dynamic" makes me think the masks will be determined on the fly based on previous measurements}\ynote{Lets use the selective sensing vacabulary here that we introduce in the title.}}: Masks can be optimized based on the tasks.
    \begin{itemize}
        \item Optical Neural Networks (ONN).
    \end{itemize}
\end{enumerate}
}

\Xhide{
\rnote{should we ignore the LFTH and STH? And add Compressed sensing}
\ynote{is this really ne distinction we need to make? Key is wether or not we change the mask to adjust to the measurement (selective sensing), or we use a general mask.}
\begin{enumerate}
    \item Null-Prior: RS, II, and HB. It requires no information before measurements.
    % \item Reconstruction-Prior: RTH and LFTH. This prior is from the inductive bias for signal reconstruction that low frequencies components are more robust to noise regardless of the data. % it is wrong if not explaining the LFTH
    \item Reconstruction-Prior: RTH. This prior is based on the fact that the light throughput and SNR in each measurement are greater with fewer masks.
    \item Task-Specific-Prior: STH, HPCA, and ONN. This prior requires the features from training data of a specific task.
    % make plots based on the groups
\end{enumerate}  
\rnote{We now have several results: classification reconstruction plot on MNIST; freeze method; reconstruction on random patterns; what to include?}  \bnote{Reconstruction on random patterns needs to be in the section when we explain the ONN optimizer}
}

\subsection{Constraints of sensing matrices} \label{ssection:constraints}

\Xpolish{Though we are able to find the ideal $\B{M}$ based on the tasks we want to accomplish, the optimization problem of $\B{M}$ may not be convex due to the following two constraints.}
{\Xpolish{While we can identify the optimal $\B{M}$ tailored to the tasks at hand, it's crucial to acknowledge that the optimization problem for $\B{M}$ might not exhibit convexity. This is attributed to the presence of two constraining factors. Consequently, achieving globally optimal masks may not be universally feasible within the scope of this problem.}{
The implementation of the sensing matrix $\B{M}$ usually involves the following constraining factors.
}
}

\begin{enumerate}
    \item\label{constraint:photonNumber} \textbf{Flux-preserving} \cite{willet2009CSPoisson}\Xhide{\bnote{Y: Should I use the same term as Rebecca?}}. The single-pixel camera model involves the allocation of available photons among masks, as discussed in \cite{neifeld2003dual}. It is important to ensure that the mask basis $\B{M}$ does not produce additional photons through improper entries \cite{neifeld2003dual}. Mathematically, $\sup \sum_{i=1}^m M_{ij} = 1, \forall j \in \{1,2,\dots,N\}$ \cite{neifeld2003dual}.\Xhide{\bnote{Include what an improper entry is!} \bnote{A: this is not something we need to mention here. its obvious that the measurements have to be normalized correctly}}    
    \item\label{constraint:negativeEntry} \textbf{Positivity-preserving} \cite{willet2009CSPoisson}. It is not possible to physically implement negative values for masks $\B{M}$, as demonstrated in \cite{neifeld2003dual, willet2009CSPoisson}. In this project, we used the dual-rail approach as outlined in the work of Neifeld et al. where the positive and negative entries of a mask give rise to two separate measurements \cite{neifeld2003dual}.
\end{enumerate}

\subsection{Model under Noise}\label{ssection:model_under_noise}
\AVnote{\checkednote{We need to better explain the noise here. Photon noise is actually super-Poissonian. it is often approximated by a Poisson distribution. The key aspect affecting the design optimization is th non-linearity of the noise. We model this nonlinearity i a quasi-classical model with a Gaussian with an intensity dependent variance.}}

\Xpolish{Two noise models were investigated in this project. The first one is the AGN, or read noise, mainly stemming from thermal vibration of atoms at sensors. In this noise model, the equation \ref{eqn:NoiselessMeasurement} turns into 
\Xequa{
    \noisy{\B{y}} = \B{Mx} + \B{\epsilon} 
},
where $\B{\epsilon} \sim \AGN(\B{0}, \sigma^2\B{I})$ and the $\sigma$ is the standard deviation. The other is the Poisson noise, or shot noise, \Xhide{model for photon-counting systems\rnote{A:shot noise doesn't just exist in photon counting systems} \cite{willet2009CSPoisson} }originating from the statistical nature of photons \cite{boyat2015review} with the following form
\Xequa{
    \noisy{\B{y}} \sim \Poisson(\B{Mx})
}. Due to the constraint \ref{constraint:negativeEntry}, no entries in $\B{M}$ can be negative.}
{This project investigates two noise models. The first one is AGN \Xpolish{or read noise}{related to dark current or read noise from imperfect sensor materials}, which mainly originates from thermal vibrations of atoms at sensors. In this noise model, Eq. \ref{eqn:NoiselessMeasurement} becomes:
\Xequa{
\label{eqn:gaussian_measurement}
\noisy{\B{y}} = \B{Mx} + \B{\epsilon},
}
 where $\B{\epsilon} \sim \AGN(\B{0}, \sigma^2\B{I})$, and $\sigma$ is the standard deviation. The other noise model is photon noise, which arises from the statistical nature of photons \cite{boyat2015review}. 
 % In a majority of previous work, it is approximated by Poisson noise and the measurement equation for this noise model usually obeys a Poisson distribution:
 {
 In most prior research, this is approximated as Poisson noise, and the measurement equations associated with this noise model typically follow a Poisson distribution:
 }
\begin{equation}
\label{eqn:poisson_measurement}
\begin{aligned}
\noisy{\B{y}} &\sim \Poisson(\B{Mx})\\
\Pr(\noisy{{y}}_i &=  k) = \frac{y_i^k \exp(-y_i)}{k!}
\end{aligned} \quad,
\end{equation}
where $y_i = \sum_{j=1}^N M_{ij} x_j$ \cite{willet2009CSPoisson}. It should be noted that the constraint \ref{constraint:negativeEntry} in Eq. \ref{eqn:poisson_measurement} prohibits negative entries in $\B{M}$.
}
% \YLnote{However, it is important to note that photon noise is inherently super-Poissonian though it is often approximated by Poissonian variables only when light-beam degeneracy is small \cite{mandel1959fluctuations}.}
% The reason that the Poisson noise is not good enough for this project is not limited to the fact that this model is not physically accurate by the context.
{
% Though Poisson noise is the best approximation of photon noise in this moment, the inadequacy of the Poisson noise model for this project extends beyond its lack of physical accuracy with regard to photon noise in this context.
% While Poisson noise serves as the most effective approximation of photon noise at this stage, its limitations for this project go beyond mere physical inaccuracies. 
While Poisson noise is the best way to approximate photon noise at this stage, its limitations for this project extend beyond just physical inaccuracies. 
}
% The non-linearity of photon noise introduces complexities in the design and optimization of systems affected, and the approximation by Poissonian variables usually gives rise to the non-existence of gradients. 
The nonlinearity of photon noise adds complexities to the design and optimization of affected systems, and the use of Poisson random functions eliminates gradients, making it difficult to optimize the hardware components in the end-to-end model.
\Xpolish{To address these shortcomings, in this project, we focus on the signal-dependency of photon noise and adopt a quasi-classical model that uses a Gaussian variable with intensity-dependent variance formulated as}{
To address these shortcomings, this project focuses on the signal dependency of photon noise and adopts a quasi-classical model known as MLGAUSS that uses a Gaussian variable with \Xpolish{intensity-dependent variance}{mean-variance equivalence} \cite{selwood2022coded_aperture_imaging}, formulated as
}
\begin{equation} \left\{
    \begin{aligned}
        \noisy{\B{y}} &= \B{Mx} + \B{J},\\
        \B{J} &=  (\B{Mx})^{\circ \frac{1}{2}} \odot \B{\epsilon},\\
        \B{\epsilon} &\sim \mathcal{N}(0, \B{I}),
    \end{aligned} \right. \quad .
\end{equation}
 \Xpolish{
 which has been introduced by Cossairt et al. and Shin et al. \cite{cossairt2012does, shin2013low} and \Xpolish{introduced}{further discussed} in section \ref{sssection:ONN Optimization under Poisson Noise}.
 }{
 \Xhide{
 This model was introduced by Cossairt et al. and Shin et al. \cite{cossairt2012does, shin2013low} and is further discussed in section \ref{sssection:ONN Optimization under Poisson Noise}.
    }
 } 
 It substitutes conventional Poisson approximations whenever a gradient computation is required, yet the term "Poisson" is retained for consistency with other classical models involving no hardware optimization, facilitating straightforward comparisons. 
 While this approach represents an alternative version of approximated photon noise besides Poisson noise, it is utilized exclusively during model training. 
 For model testing, all models employ the Poisson noise model, as gradient computation is no longer necessary.
 \Xhide{Although there is an inconsistency between the noise models used for training and testing, this discrepancy is not significant as long as we adhere to the signal-dependency property, which allows for optimal or near-optimal solutions.
 Supporting this perspective, the simulated and experimental results demonstrate consistent performance improvement across both stages.}

\Xhide{
Two noise models were investigated in this project. The first one is the 
% Additive White Gaussian Noise (AWGN)
AGN
, or read noise,  mainly stemming from thermal vibration of atoms at sensors \iffalse the imperfectness of sensors\fi \cite{boyat2015review} with the following form
% \subsubsection{Gaussian Noise Model}
\Xequa{
    \Xalign{
        \lambda\tilde{\B{y}^+} &\sim \mathcal{N}(\lambda \B{M}^+ \B{x} , \sigma^2 \B{I})\\
        \lambda\tilde{\B{y}^-} &\sim \mathcal{N}(\lambda \B{M}^- \B{x} , \sigma^2 \B{I})
    },
} 
% \subsubsection{Poisson Noise Model}
where the $\sigma$ is the standard deviation. The other is the Poisson noise, or shot noise, \Xhide{model for photon-counting systems\rnote{A:shot noise doesn't just exist in photon counting systems} \cite{willet2009CSPoisson} }originating from the statistical nature of photons \cite{boyat2015review} with the following form
\Xequa{
    \Xalign{
        \lambda\tilde{\B{y}^+} &\sim \mathcal{P}(\lambda \B{M}^+ \B{x})\\
        \lambda\tilde{\B{y}^-} &\sim \mathcal{P}(\lambda \B{M}^- \B{x})
    }.
}   
\bnote{This paragraph is somewhat confusing, and some of the ideas should maybe be before the equations.}  
Although we can obtain a negative number of photons from the difference between the two branches, the Poisson noise cannot be directly applied to these values. Otherwise, it violates the nature of Poisson distributions and the fact that Poisson noise appears in the sensors. On the contrary, noise should be considered in both branches independently.  
\bnote{Mention that this comes from the relevant equation in the dual branch section. }
Therefore, in the AGN model, the measurement $\lambda \tilde{\B{y}} \sim \mathcal{N}(\lambda \B{Mx}, 2\sigma^2 \B{I})$ with the dual-rail trick. But in the Poisson noise model, the measurement $\lambda \tilde{\B{y}} \sim \text{Skellam}(\lambda \B{M}^+ \B{x}, \lambda \B{M}^- \B{x})$ \cite{hwang2007skellam, gan2017skellam}. 
}

% \subsection{Scan Strategies with Different Priors}

% \bnote{We should maybe focus less on the term "single-pixel imaging" that is the way we show our results but maybe it obscures our main point.}

% \Xhide{\ynote{We already talked about single pixel imaging in the beginning. I think the figure could be introduced here along with the math model}
% Single-pixel imaging is a technique that involves both computational and non-computational methods. }

\section{Methods}
\label{sec:4_methods}
\subsection{Model Configuration \ynote{\checkednote{this is just the ONN. We probably also want to say something about PCA}}} 
\begin{figure}[ht]
    \centering
    \begin{subfigure}{\linewidth}
    \caption{Scanner-Classifier Configuration}
    % \includesvg[width=0.9\linewidth]{\homedir/figures/Methods/Model_Structure.svg}
    \includegraphics[width=0.9\linewidth]{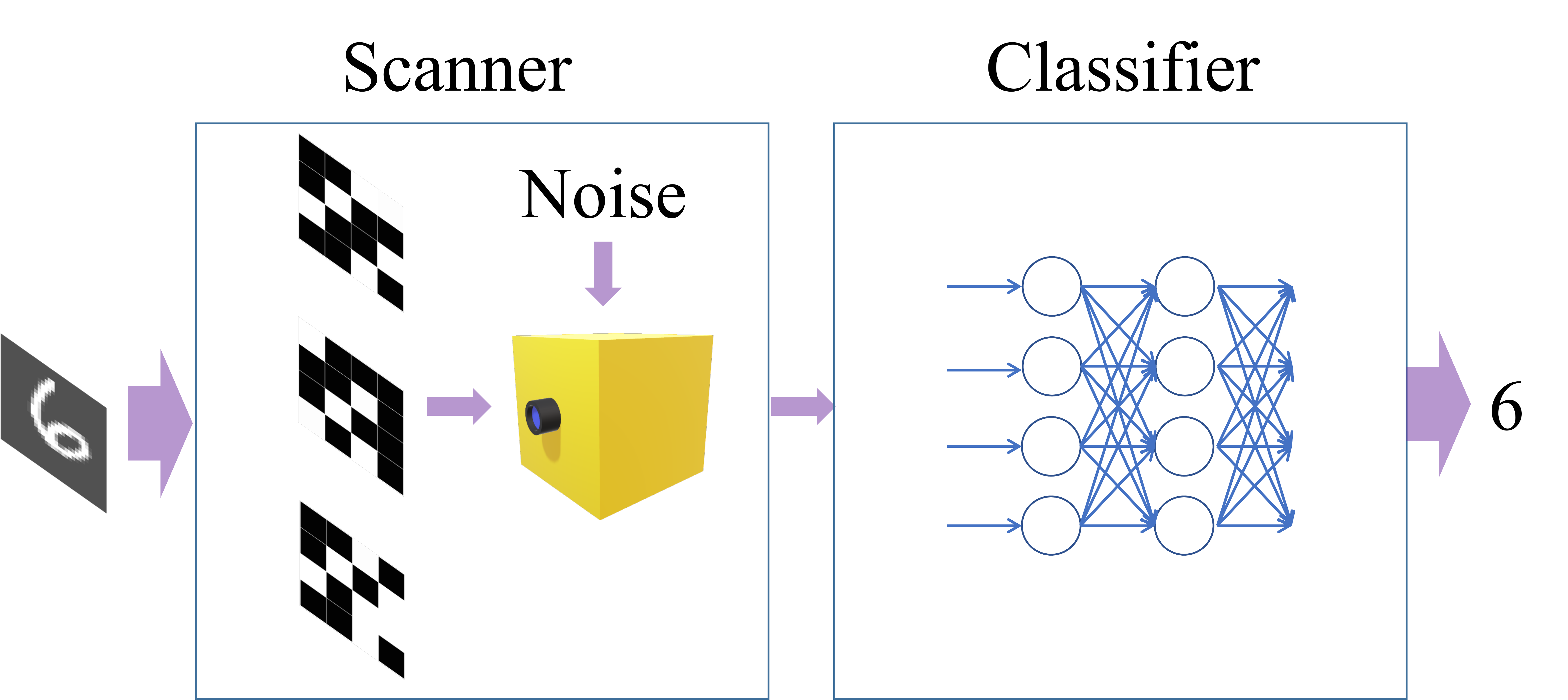}    
    \label{fig:model_config_a}
    \end{subfigure}

    \begin{subfigure}{\linewidth}    
    \caption{Graph Representation}
    % \documentclass{article}
% \usepackage{tikz}
% \usetikzlibrary{shapes.geometric}
% \usepackage{caption}
% \usepackage{amsmath,amsfonts}

\newcommand{\Wd}{0.14\linewidth}
\newcommand{\Ceil}{1.8}
\newcommand{\Floor}{-1.8}
\newcommand{\Ht}{0}

% \begin{document}

\tikzset{%
  every input_neuron/.style={
    circle,
    draw = black
    },    
  every optical_neuron/.style={
    % rectangle,
    % chamfered rectangle,
    % regular polygon,
    diamond,
    fill=green,
    draw = black,
    minimum size=0.8cm
    },
  every neuron/.style={
    circle,
    draw,
    minimum size=0.5cm,
    fill=cyan
  },
  neuron missing/.style={
    draw=none, 
    fill=none,
    scale=1,
    % text height=0.1cm,
    text height=0.5cm,
    execute at begin node=\color{black}$\vdots$
  },
  every noise/.style={
    shape=diamond,
    fill=green,
    draw=black,
    minimum size=0.8cm
  }
}

% \begin{figure}[ht]

% \centering

\begin{tikzpicture}[x=1cm, y=1cm, >=stealth]

%如果有4个包括省略号: {1,2,missing,3,4}
% \foreach \m/\l [count=\y from 0] in {1,2,3,missing,4,5,6}
%   \node [every input_neuron/.try, neuron \m/.try] (input-\m) at (0*\Wd,\Ceil - \y * \Ceil / 6 + \y * \Floor / 6) {};
\foreach \m/\l [count=\y from 0] in {1,2,missing,3,4}
  \node [every input_neuron/.try, neuron \m/.try] (input-\m) at (0*\Wd,\Ceil - \y * \Ceil / 4 + \y * \Floor / 4) {};
  
\foreach \m [count=\y from 0] in {1,2,missing,3,4}
  \node [every optical_neuron/.try, neuron \m/.try] (DMD-\m) at (1*\Wd,\Ceil - \y * \Ceil / 4 + \y * \Floor / 4) {};

\foreach \m [count=\y from 0] in {1,2,missing,3,4}
  \node [every noise/.try, neuron \m/.try] (noise-\m) at (2*\Wd,\Ceil - \y * \Ceil / 4 + \y * \Floor / 4) {};
  
\foreach \m [count=\y from 0] in {1,2,missing,3,4}
  \node [every neuron/.try, neuron \m/.try ] (hidden1-\m) at (3*\Wd,\Ceil - \y * \Ceil / 4 + \y * \Floor / 4) {};

\foreach \m [count=\y from 0] in {1,2,missing,3,4}
  \node [every neuron/.try, neuron \m/.try ] (hidden2-\m) at (4*\Wd,\Ceil - \y * \Ceil / 4 + \y * \Floor / 4) {};

% \foreach \m [count=\y from 0] in {1,2}
%   \node [every neuron/.try, neuron \m/.try ] (output-\m) at (4*\Wd,\Ceil/2 - \y * \Ceil / 2 + \y * \Floor / 2) {};

\foreach \m [count=\y from 0] in {1,2,missing,3,4}
  \node [every neuron/.try, neuron \m/.try ] (output-\m) at (5*\Wd,\Ceil - \y * \Ceil / 4 + \y * \Floor / 4) {};

%%%%%%%%%%%

% \foreach \l [count=\i] in {1,2,3,n-2,n-1,n}
\foreach \l [count=\i] in {1,2,n-1,n}
  \draw [<-] (input-\i) -- ++(-1,0)
    node [above, midway] {$X_{\l}$};

\foreach \l [count=\i] in {1,2,m-1,m}
  \node at (DMD-\i) {\textcolor{red}{\tiny${\l}$}};
\foreach \l [count=\i] in {1,2,m-1,m}
  \node at (noise-\i) {\textcolor{red}{\tiny${\l}$}};
\foreach \l [count=\i] in {1,2,39,40}
  \node at (hidden1-\i) {\tiny${\l}$};
% \foreach \l [count=\i] in {1,...,4}
\foreach \l [count=\i] in {1,2,127,128}
  \node at (hidden2-\i) {\tiny${\l}$};
  
\iffalse
\foreach \l [count=\i] in {1,2,7,8}
  \node [above] at (hidden1-\i.north) {$H_{\l}^{(1)}$};
  
\foreach \l [count=\i] in {1,...,4}
  \node [above] at (hidden2-\i.north) {$H_{\l}^{(2)}$};
\fi
 
%\foreach \l [count=\i] in {1,...,2}
%  \node [above] at (output-\i.north) {$O_{\l}$};
    
\foreach \l [count=\i] in {1,2,9,10} % output
  \draw [->] (output-\i) -- ++(1,0)
    node [above, midway] {$O_{\l}$};

% \foreach \i in {1,...,6}
\foreach \i in {1,...,4}
  \foreach \j in {1,...,4}
    \draw [->] (input-\i) -- (DMD-\j);
    
\foreach \i in {1,...,4}
    \draw [->] (DMD-\i) -- (noise-\i);

\foreach \i in {1,...,4}
  \foreach \j in {1,...,4}
    \draw [->] (noise-\i) -- (hidden1-\j);

\foreach \i in {1,...,4}
  \foreach \j in {1,...,4}
    \draw [->] (hidden1-\i) -- (hidden2-\j);

\foreach \i in {1,...,4}
  \foreach \j in {1,...,4}
    \draw [->] (hidden2-\i) -- (output-\j);

\foreach \l [count=\x from 0] in {Input, $\B{M}$, Noise,  ,  , Output}
  % \node [align=center, above] at (\x*\Wd,\Ceil+1) {\l \\ layer};
  \node [align=center, above] at (\x*\Wd,\Ceil+0.5) {\l};

  \node [align=center, above] at (3.5*\Wd,\Ceil+0.5) {Fully connected \\layers};

\end{tikzpicture}

% \captionsetup{labelformat=empty}

% \end{figure}

% \end{document}
    \label{fig:model_config_b}
    \end{subfigure}
    \caption{(\subref{fig:model_config_a}) the general configuration of scanner-classifier networks. (\subref{fig:model_config_b}) the node-level architecture of our scanner (green diamonds)-classifier (blue circles) network. \Xhide{The input $X_i$ is a handwritten number and the output $O_j$ is the probability that $X_i$ is number $j$. }Noise is implemented after the sensing matrix $\B{M}$. Different from other coding schemes, ONN has a trainable scanner where $\B{M}$ is not fixed and can be optimized by the gradient of the classifier.}
    \label{fig:model_config}
    % \label{fig:ONN_architecture}
\end{figure}

\Xpolish{Although coding strategies are different, they essentially share the same model structure shown in figure \ref{fig:model_config} for classification tasks. In this study, we built a classification pipeline consisting of a scanner module and a classifier in PyTorch. This model consists of a \textit{scanner} module and a \textit{classifier} module.\Xhide{The scanner is simulating the measurement process described in equation \ref{eqn:NoisyMeasurement} and the classifier is a neural networks model.} The scanner simulates the operations to count the photons, following the techniques\Xhide{ \bnote{"tricks" -> "techniques"}} mentioned in section \ref{ssec:ModelFormulation}. We\Xhide{ also \bnote{"Also" could mean we have two different models.}} employed a two-hidden layer artificial neural network (ANN) classifier with 40 and 128 nodes as the main model for predicting which number the object is.
%by the corresponding  number of re-centered, normalized, and noisy photon counts. 
% The input data for the classifier was obtained by subtracting the average of the training data, represented by $\B{\bar{x}}_T$, from the re-centered normalized noisy photon counts, denoted by $\tilde{\B{y}}$. 
To prevent overfitting and improve the generalizability of the model, dropout  was applied after each hidden layer, followed by a ReLU activation layer.  }
{All coding strategies in this project for classification tasks share the same fundamental model structure, as illustrated in Fig. \ref{fig:model_config}\Xhide{ and \ref{fig:ONN_architecture}}. \Xpolish{In this project, we extend the applicability of our model to three distinct tasks: image reconstruction\AVnote{This idea, which focuses on image reconstruction, is based on prior knowledge, such as signals being sparse in some linear space, with noise concentrated in the high-frequency domain. One example in this category is compressed sensing. Apr 10, 2024 12:38 PM}, handwritten number classification with the MNIST dataset, and classification of hyperspectral data.}{
\YLnote{In this project, we apply our model to handwritten number classification with the MNIST dataset and classification of hyperspectral data.}
} To classify handwritten numbers using optical devices with limited photon budget, we built a PyTorch-based pipeline consisting of a \textit{scanner} module \YLnote{including $\B{M}$ and a noise generator} to simulate the acquisition of $\noisy{\B{y}}$ described in Eq. \ref{eqn:gaussian_measurement} and Eq. \ref{eqn:poisson_measurement}, and a \textit{classifier} module classifying the measured objects by $\noisy{\B{y}}$.\Xhide{ The scanner module simulates the photon counting process using the techniques described in section \ref{ssec:selectionofsensingmatrix}.} For the classification task, we utilized a two-hidden layer artificial neural network (ANN) with 40 and 128 nodes, serving as the primary model. To prevent overfitting and enhance generalizability, we applied dropout after each hidden layer, followed by a rectified linear unit (ReLU) activation layer.}
Once operational, this model can serve various purposes, such as object classification and signal recovery. For instance, it can process images of handwritten numbers, providing the likelihood of each image corresponding to a specific number by cross-entropy loss. 
Furthermore, the classifier can function as a decoder to recover the input image by MSE loss, which is adopted in section \ref{ssec:model_validation}.

\Xpolish{When training this model, the scanner computes the photon counts with noise, and feeds it to the classifier. The noise is sampled from a predefined noise distribution and is regenerated in different epochs. After the loss is computed, the parameters in the classifier are optimized by gradient descent. The ONN is different from the others as the back-propagation is extended into the scanner so that the masks are optimized together with the classifier.}{
During model training, the scanner computes the photon counts with noise and then transmits these data to the classifier. 
The noise is sampled from a predefined noise distribution and varies across epochs. Once the loss has been computed, the gradient descent is used to optimize the classifier parameters. 
\YLnote{In addition, we propose an end-to-end optimization model called Optical Neural Networks (ONN).} 
\Xhide{\textbf}{What sets the ONN apart from other \Xpolish{models}{coding schemes} is the learnable $\B{M}$ which is not predefined or fixed. 
In other words, backpropagation is extended to the scanner, allowing the simultaneous optimization of the masks and the classifier under different noise models}. 
This approach offers distinct advantages compared to traditional methods, which typically optimize the classifier independently of the scanner or optimize the scanner without the appropriate noise model.}

\YLnote{\checkednote{Another figure showing network}}

\Xpolish{ONN can also be used for reconstruction. \YLnote{To address this, we propose a simple approach of optimizing the ONN masks for a reconstruction task under different noise models. To achieve this, we developed a new ONN model with $m = N$ masks. The scanner takes raw image vectors $\B{x}$ and outputs $N$-element measured photon count vectors under a chosen noise model, $\noisy{\B{y}}$. The classifier is a single-layer network that takes $\noisy{\B{y}}$ as input and outputs $\noisy{\B{x}}$, also an $N$-element vector. The objective is to minimize the MSE between the input images and the output vectors, $\|\noisy{\B{x}} - \B{x}\|_2^2$. In noiseless cases, the classifier is equivalent to $\B{M}^{-1}$, where $\B{M}$ is the masks used in the scanner.}}{}

\subsubsection*{ONN Optimization under \Xpolish{Photon}{Poisson} Noise} 
\label{sssection:ONN Optimization under Poisson Noise}

\Xpolish{Since the Poisson noise generating function has no gradient, it is challenging to optimize the scanner. The reparameterization trick is a useful method for the AGN instead of the Poisson noise. Our method is using a normal distribution whose variance equals its expectation to approximate the Poisson distribution in training, and therefore, the reprameterization trick can be employed.}{
Optimizing the scanner can be challenging, as the Poisson \Xpolish{noise generating}{random} function typically lacks a gradient. 
% While reparameterization is a useful method for addressing AGN, it is not as effective for addressing Poisson noise. To address this limitation, we employ a method mentioned by Cossairt et al. and Shin et al. that utilizes a normal distribution with a variance equal to its expectation \Xpolish{to approximate}{instead of} a Poisson distribution during training \cite{cossairt2012does, shin2013low}. 
We employ the reparameterized-Gaussian model mentioned in section \ref{ssection:model_under_noise} to address this limitation. 
\ynote{\checkednote{we need more details on this and how it is implemented in the network. Probably a figure of the entire network structure and the equation for the gradient (i think we also have references to cite here)}}
This \Xpolish{approximation}{approach} enables us to employ the PyTorch built-in reparameterization method, \Xpolish{i.e.}{that is,} \texttt{rsample}, and optimize the scanner more effectively under \Xpolish{photon}{Poisson} noise.}
% \Xpolish{During the training under \Xpolish{Poisson} noise}{In the training stages}, the variance also plays a role in gradient computation, which is the most significant difference compared with training \Xpolish{under AGN}{under conventional Poisson models}.
During the training stages, variance plays a crucial role in the gradient computation, marking a significant difference compared to training under conventional Poisson models.

% \YLnote{
% A seperate section? And mention that Ollie did that as well.  \\
% Similar to corssairt et al, we use a normal dist to approximate poisson
% }

% \YLnote{
% Transform $\B{y} \sim \mathcal{P}(\B{Mx})$ to $\B{y} \sim \mathcal{N}(\B{Mx}, (\B{Mx})^\frac{1}{2})$ to use the reparamaterization trick. The variance will also play a role in computing the gradient.
% } 

% To approximate \Xpolish{Poisson}{photon} noise by AGN, we can re-write the measurement with noise $\noisy{\B{y}} = \B{Mx} + \B{J}$ where $\B{J} =  (\B{Mx})^{\circ \frac{1}{2}} \odot \B{\epsilon}, \B{\epsilon} \sim \mathcal{N}(0, \B{I})$ \cite{cossairt2012does, shin2013low}. During the training under \Xpolish{Poisson}{photon} noise, $\B{J}$ also plays a role in gradient computation, which is the most significant difference compared with training under AGN. 

\textcolor{blue}{\checkednote{I think we can remove sections 4.2 and 4.3. They mostly repeat waht we already said and add too much detail for a paper.}}

\subsection{Model Validation}
\label{ssec:model_validation}
% In this section, we start with a theoretical investigation on the reconstruction performances of null-prior strategies. Then the reconstruction and classification performances of all strategies and fairly compared by simulations and experiments. 

\YLnote{
{
To validate the proposed ONN model, we apply it to non-compressible signal reconstruction tasks where the correct solution is known. \Xpolish{In subsequent sections, we derive the theoretical error propagation caused by both AGN and Poisson noise, assessing the ONN model's ability to substantiate our findings.}{
Previous theoretical studies have shown that multiplexing, such as Hadamard coding, is not recommended under Poisson noise for greater MSE \cite{duarte2008CS} and lower SNR \cite{harwit1979hadamard} compared to raster scan. In this project, we compare Hadamard and raster coding under both AGN and Poisson noise conditions to see whether the ONN model reaches similar conclusions.
}
}
}
\Xhide{\AVnote{\checkednote{This section still doesn't clearly state what we do and why. Can we just say that we train masks to reconstruct non -compressible signals? Jan 18, 2024 10:58 AM}}
\AVnote{\checkednote{Need to be more specific what we show. I think we select two patterns, raster and hadamart, and prove which is better for non-compressible data. Then we see if the ONN finds similar masks. Jan 18, 2024 11:08 AM}}}

% \subsubsection*{ONN Non-compressible Signal Reconstruction}  

{ 
Given the universality of reconstruction algorithms, it is imperative that the input data vector $\B{x}$, is not compressible.
\Xpolish{
{To \Xpolish{make this reconstruction task less specific and more general}{generate non-compressible data}, we reshape the input data vectors $\B{x}$ from $10^5$ random 8 by 8 patterns, increasing their generality. 
W}{Therefore, w}e randomly sample \YLnote{\Xhide{10 times}}from the uniform distribution $U(0,1)$ to generate \Xhide{10 groups of }inputs \YLnote{$\B{x}$} and it is regenerated several times during the optimization, introducing \YLnote{\Xpolish{additional variability \AVnote{added to what?}}{extra randomization and less compressibility}}}{
\YLnote{
Hence, we generate {$\B{x}$} by randomly sampling from the uniform distribution $U(0,1)$. Throughout the optimization process, these inputs are regenerated 10 times, thereby introducing additional randomization and reducing compressibility}}. 
The output is the reconstructed $\B{x}$, with MSE between the input and the output serving as the loss function.
The simulations are conducted \Xhide{under \YLnote{\Xpolish{significant noise}{a low-light condition}}, }with a total of $1 \times 10^2$ photons which is used to rescale the $\B{x}$ to simulate a short exposure, given that AGN and Poisson noise behave more differently under low-light conditions. We initialize the masks $\B{M}$ with an $N \times N$ identity matrix and a $\{0,1\}$ Hadamard matrix, respectively, and optimize them until the model reports constant losses.
}

\Xpolish{The first 6 masks optimized by the ONN are shown in figure \ref{fig:recon_masks}. From it, we can see that the ONN model has very different behaviors under the AGN and the Poisson noise. For the AGN, it tended to open more pixels to increase the light throughput, while it adhered to the RS mode for the Poisson noise. The figure \ref{fig:masks_value_dist} gives a more thorough visualization of the optimized masks by showing their entries' distributions. The x-axis is the values in the masks and the y-axis is the total number of entries in a certain range. This result is consistent with the classification results in the last section, where the RS is mediocre under the AGN but much better under the Poisson noise.}{
Fig. \ref{fig:recon_masks} displays the first 6 optimized masks \YLnote{from both initial matrices} generated by the ONN model \YLnote{\Xpolish{. A notable observation is that the model exhibits different behavior patterns under the AGN and Poisson noise}{under both AGN and Poisson noise}}. 
Specifically, it tends to open more pixels to enhance light throughput under AGN, but adheres to the RS mode under Poisson noise. 
Fig. \ref{fig:masks_value_dist} provides a more detailed visualization of the optimized masks by presenting the distribution of their entries. 
The x-axis represents the values in the masks, while the y-axis shows the total number of entries within a specific range. 
This observation aligns with the reconstruction results\Xhide{classification results presented in the previous section}, where the RS performed modestly under AGN but considerably better under Poisson noise. 
\Xpolish{The rationale behind this is that re-projection from a captured basis into a basis of sparsity does not yield the same recovery quality under Poisson noise that is provides for AGN. To obtain a performance benefit over the trivial point scanning method, or RS, it is essential that the data is sparse and is captured in a sparse basis. Since random patterns are not sparse, the best scanning strategy is point scanning, which matches our results.}{}  
Gradients incorporating these two noise models during the training process lead mask optimization in very different directions. 
From this example, we can see that it is not appropriate to design a coding system under AGN and then apply it to Poisson noise data.
}

%diagbox example
% \begin{table}[h!]
% \centering
% \begin{tabular}{|c|c|}
% \hline
% Cell 1 & \diagbox{Top-right}{Bottom-left} \\
% \hline
% Cell 2 & Cell 3 \\
% \hline
% \end{tabular}
% \caption{Table with a diagonal split.}
% \end{table}

% \begin{table}[h!]
% \centering
% \begin{tabular}{|c|c|}
% \hline
% Cell 1 & \parbox[c][3cm][t]{3cm}{\raggedleft Top-right text \\[1cm] \raggedright Bottom-left text} \\
% \hline
% Cell 2 & Cell 3 \\
% \hline
% \end{tabular}
% \caption{Table with a cell split into bottom-left and top-right.}
% \end{table}

\begin{figure}[ht]  
\begin{subfigure}{1\linewidth}
% \caption{First 6 ONN Optimized Masks}
\caption{6 Random Selected ONN Optimized Masks}
\footnotesize	
\centering
\begin{tabular}{>{\centering\arraybackslash}p{0.15\linewidth}|>{\centering\arraybackslash}m{0.35\linewidth}|>{\centering\arraybackslash}m{0.35\linewidth}}
% {Initial matrix} &  Gaussian & Poisson\\
{}  &  Gaussian & Poisson\\
\hline
Identity & \includegraphics[width=1 \linewidth]{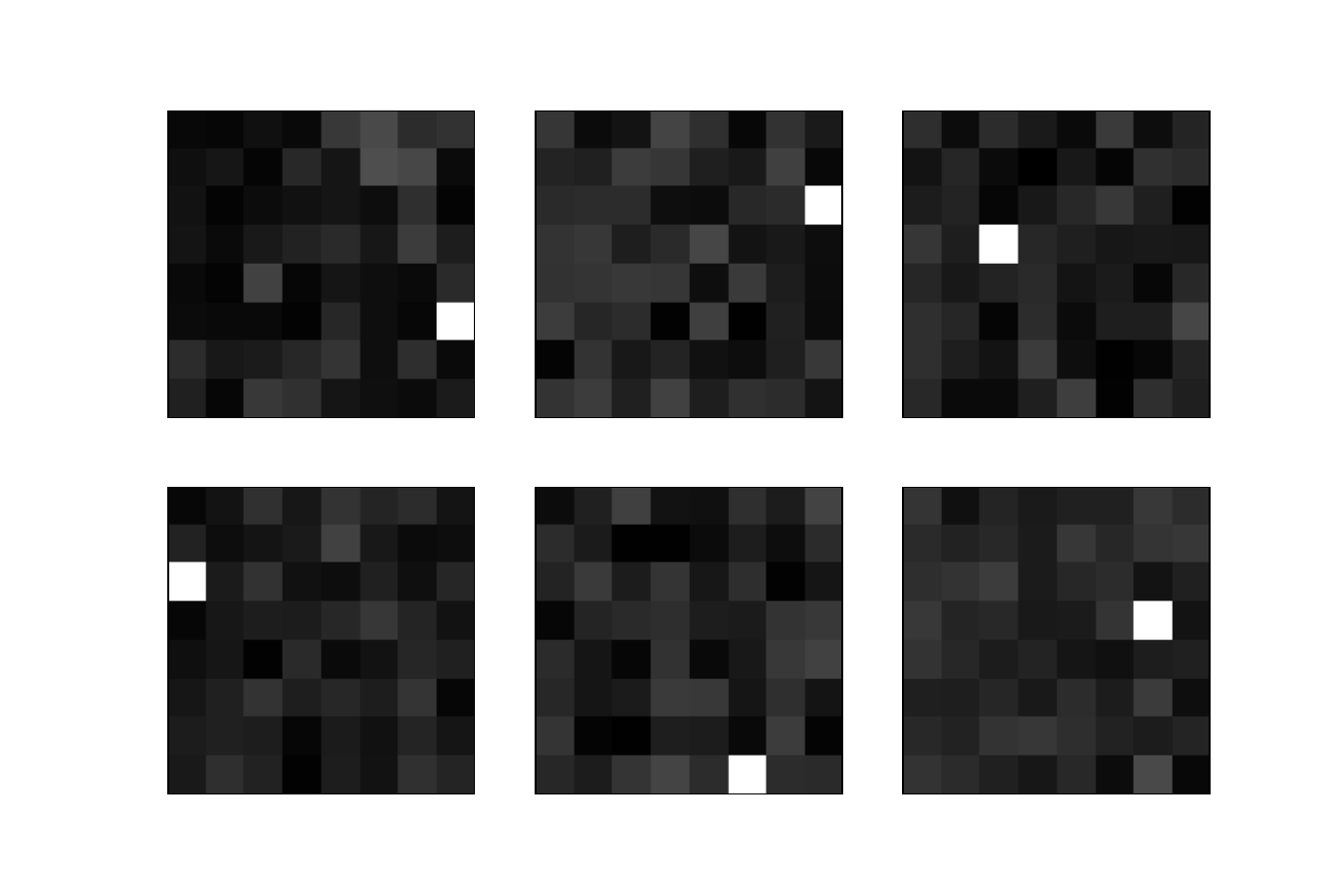} & 
\includegraphics[width=1\linewidth]{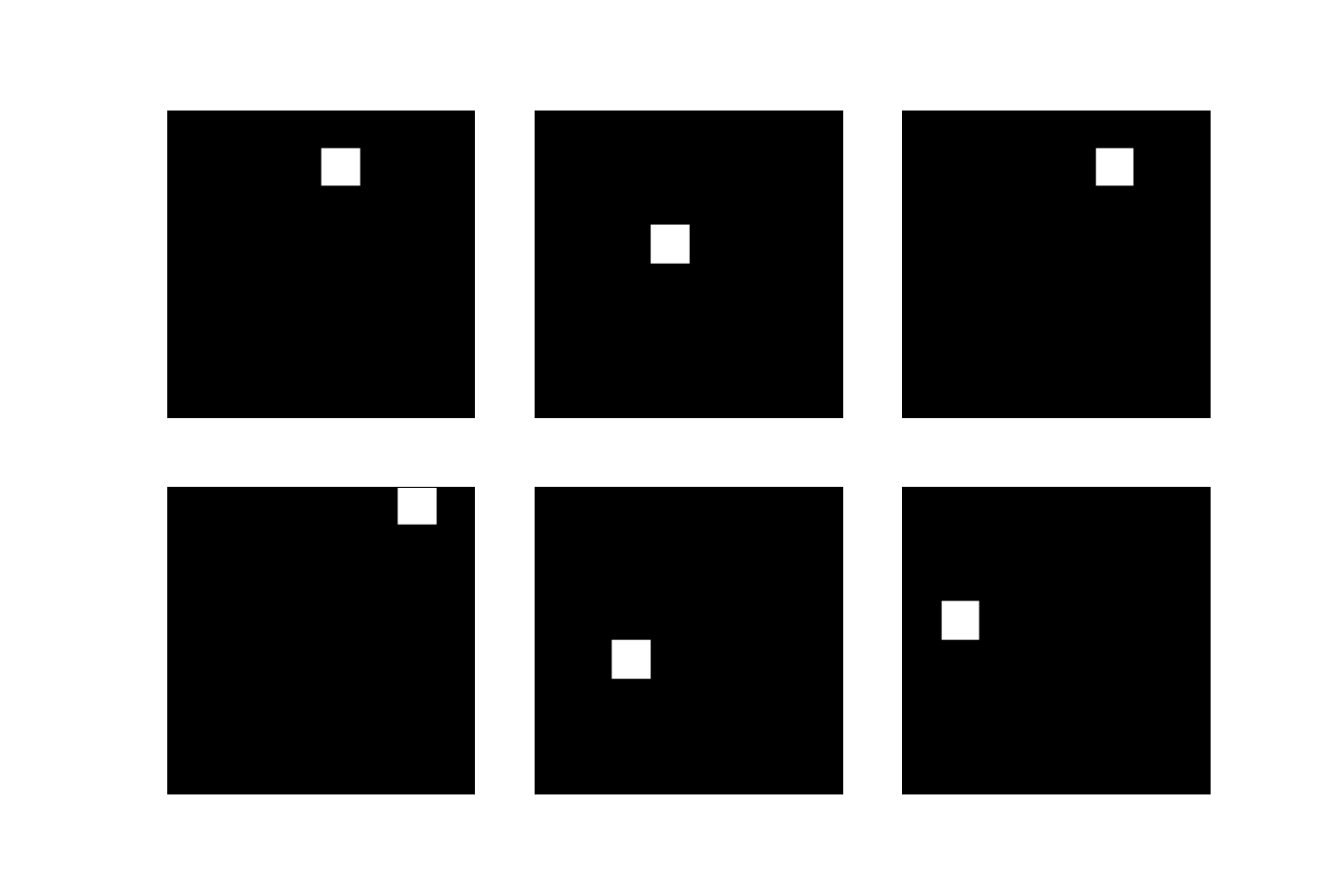} \\
\hline
Hadamard & \includegraphics[width=1 \linewidth]{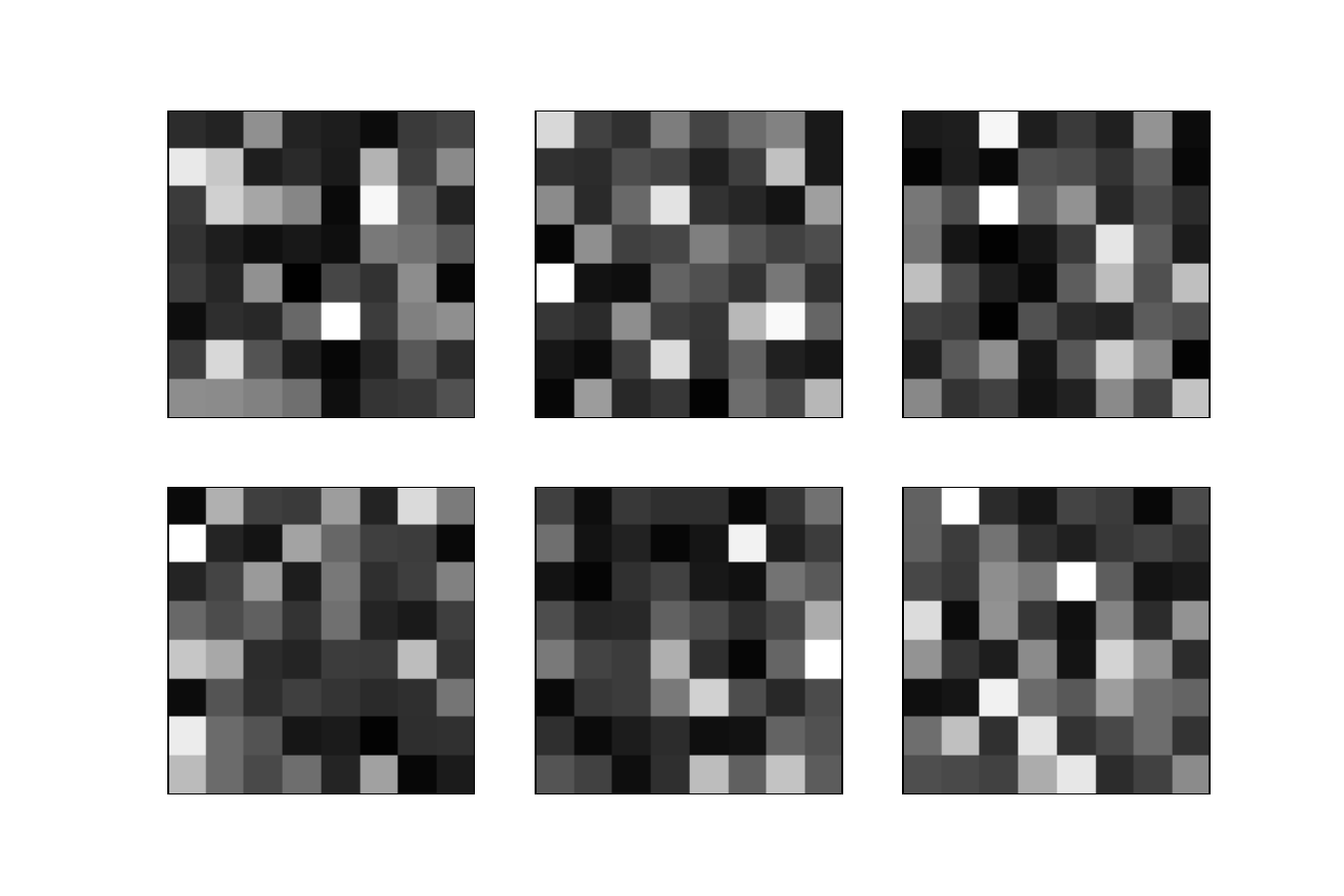} & 
\includegraphics[width=1\linewidth]{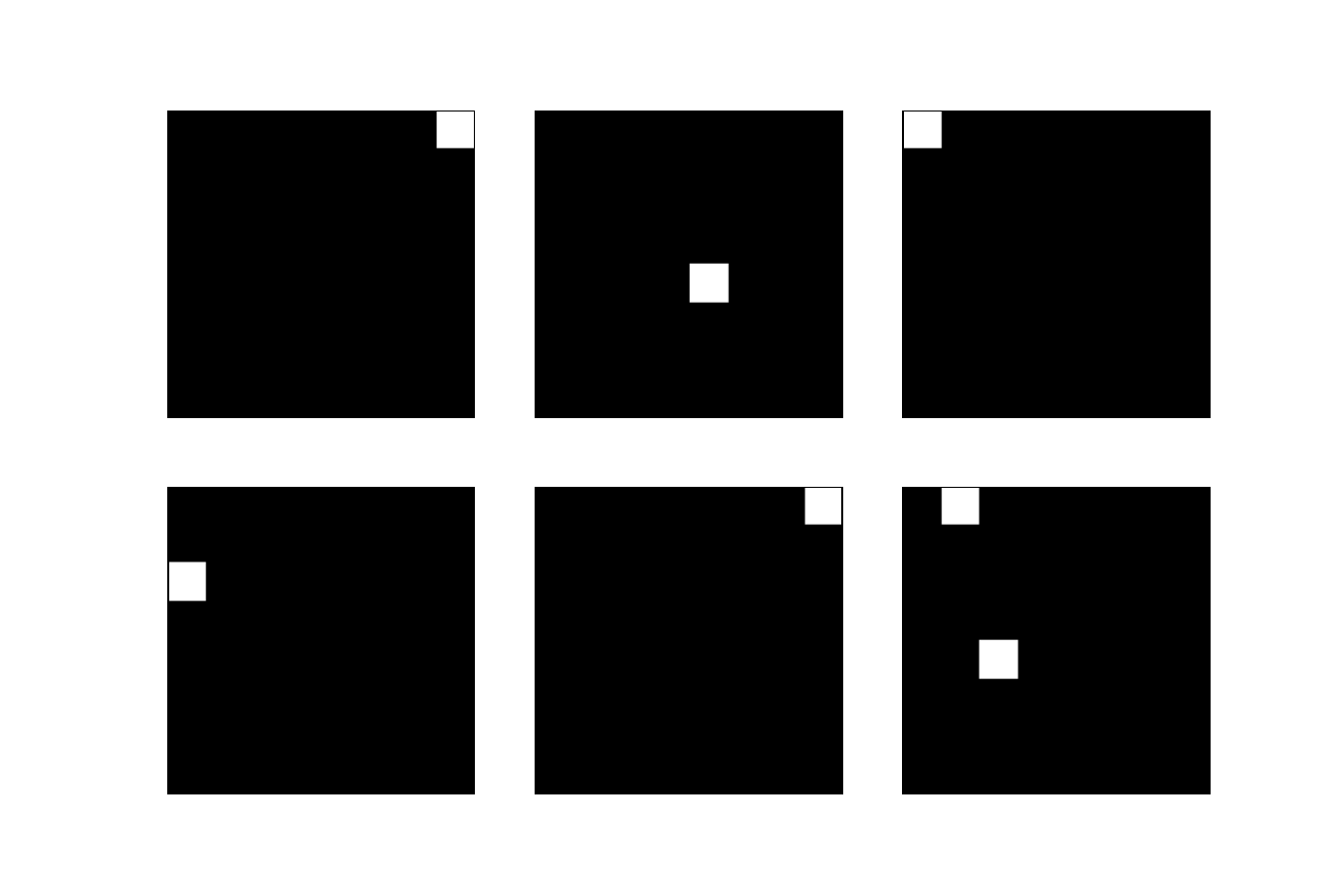} \\
\end{tabular}
\normalsize

\label{fig:recon_masks}
\end{subfigure}
\begin{subfigure}{1\linewidth}
\caption{Value Distributions of ONN Optimized Masks}
\footnotesize
\centering
\begin{tabular}{>{\centering\arraybackslash}p{0.15\linewidth}|>{\centering\arraybackslash}m{0.35\linewidth}|>{\centering\arraybackslash}m{0.35\linewidth}}
% {Initial matrix}  &  Gaussian & Poisson\\
{}  &  Gaussian & Poisson\\
 % \parbox[c][0.1\linewidth][t]{\linewidth}{\raggedleft Noise \\[0.1\linewidth] \raggedright Initial matrix} & Gaussian & Poisson \\
 % \diagbox[width=0.5\linewidth, height=5pt]{\scriptsize Noise}{\scriptsize Initial}  & Gaussian & Poisson
\hline
Identity & \includegraphics[width=1 \linewidth]{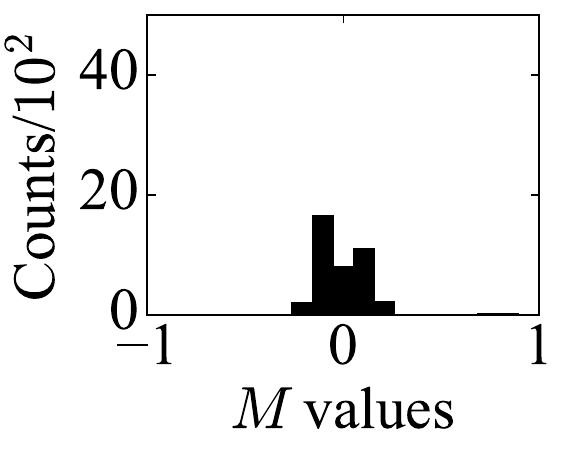} & 
\includegraphics[width=1\linewidth]{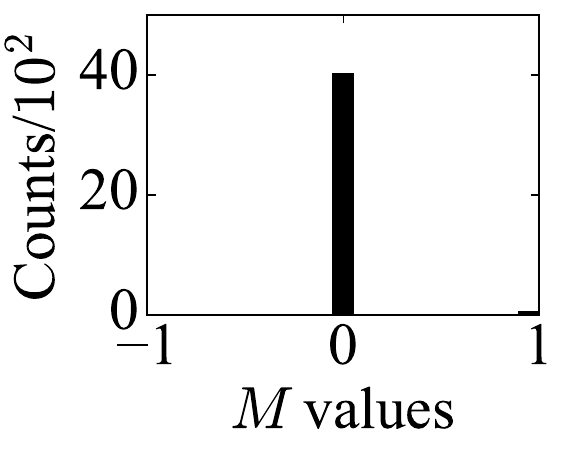} \\
\hline
Hadamard & \includegraphics[width=1 \linewidth]{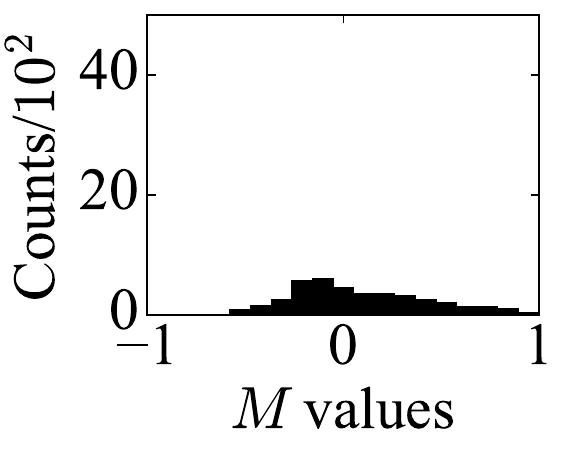} & 
\includegraphics[width=1\linewidth]{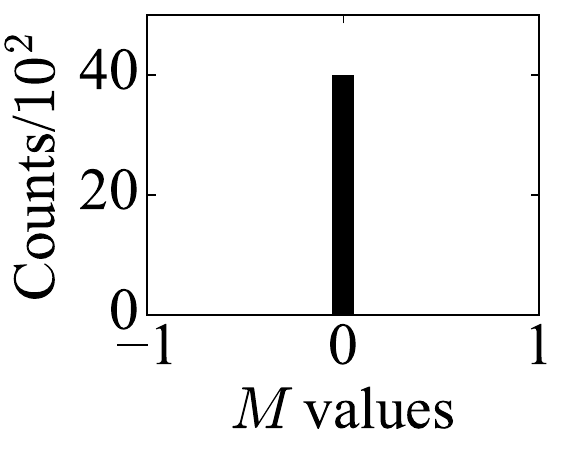} \\
\end{tabular}
\normalsize
\label{fig:masks_value_dist}
\end{subfigure}
\caption{\Xpolish{Optimizations on $\B{M}$ initialized with identity and Hadamard matrices are studied under the AGN and Poisson noise. $\B{M}$ prefers coding strategies under the AGN but favors the raster scan strategy under the Poisson noise regardless of the initialized matrices.}
{
\Xhide{(\subref{fig:recon_masks}) }Optimizations on the matrix $\B{M}$, initialized with identity and Hadamard matrices, are explored under both AGN and Poisson noise. (\subref{fig:recon_masks}) shows some optimized masks and (\subref{fig:masks_value_dist}) shows the distributions of all entries in $\B{M}$ after optimization. While $\B{M}$ demonstrates a preference for strategies with higher light throuput under AGN, it consistently leans toward the raster scan strategy under Poisson noise, irrespective of the initialization matrices. The challenge arises from the non-convex nature of this optimization problem, making it difficult to converge perfectly to the identity matrix. Nevertheless, this deviation underscores fundamental distinctions in optimizing masks under these distinct noise models.
}}
\end{figure}

\Xhide{
\begin{figure}[ht]
\caption*{First 6 Optimized Masks}
\centering
\begin{tabular}{>{\centering\arraybackslash}p{0.15\linewidth}|>{\centering\arraybackslash}m{0.35\linewidth}|>{\centering\arraybackslash}m{0.35\linewidth}}
\Xhide{Initial matrix} &  Gaussian & Poisson\\
\hline
Identity & \includegraphics[width=1 \linewidth]{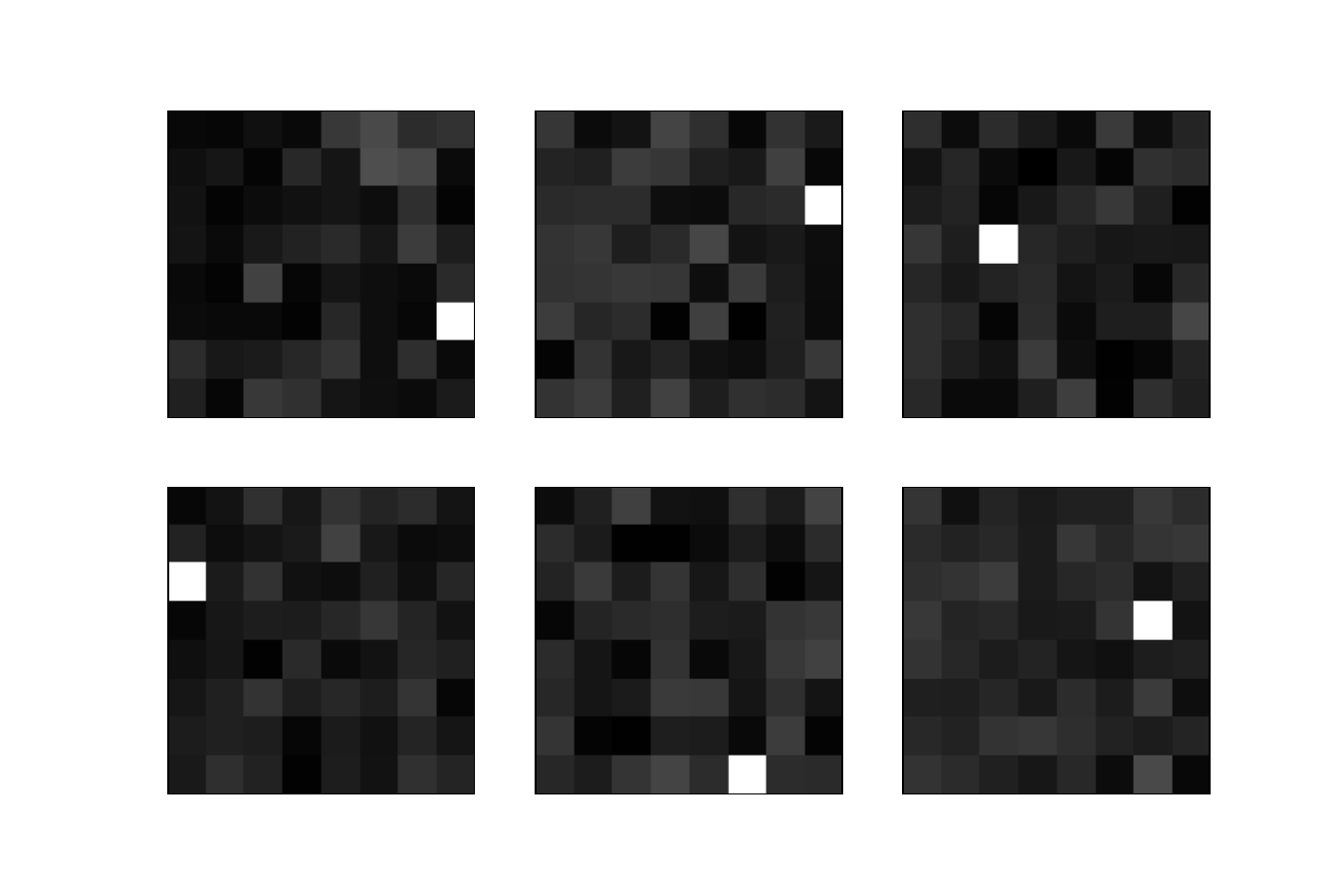} & 
\includegraphics[width=1\linewidth]{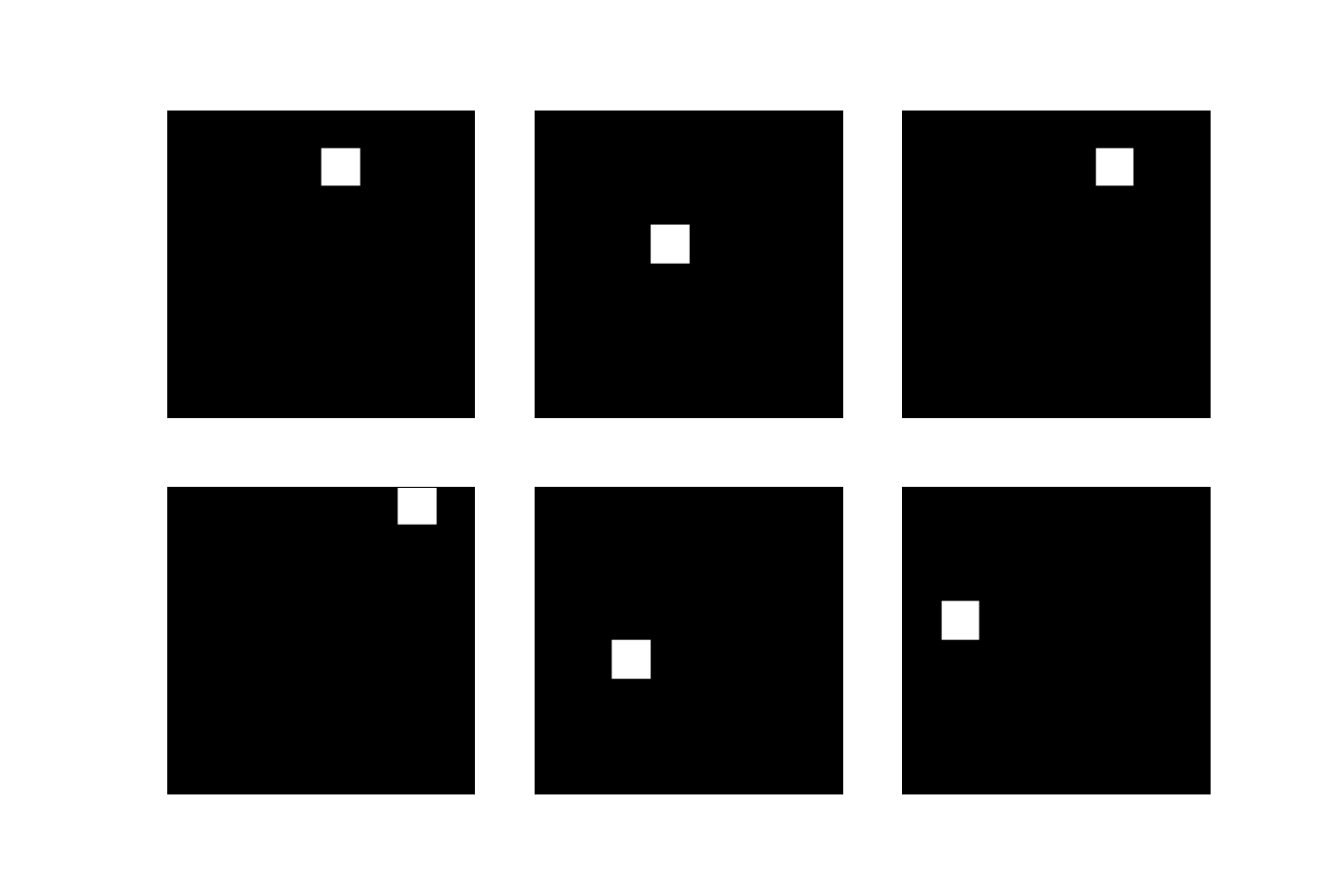} \\
\hline
Hadamard & \includegraphics[width=1 \linewidth]{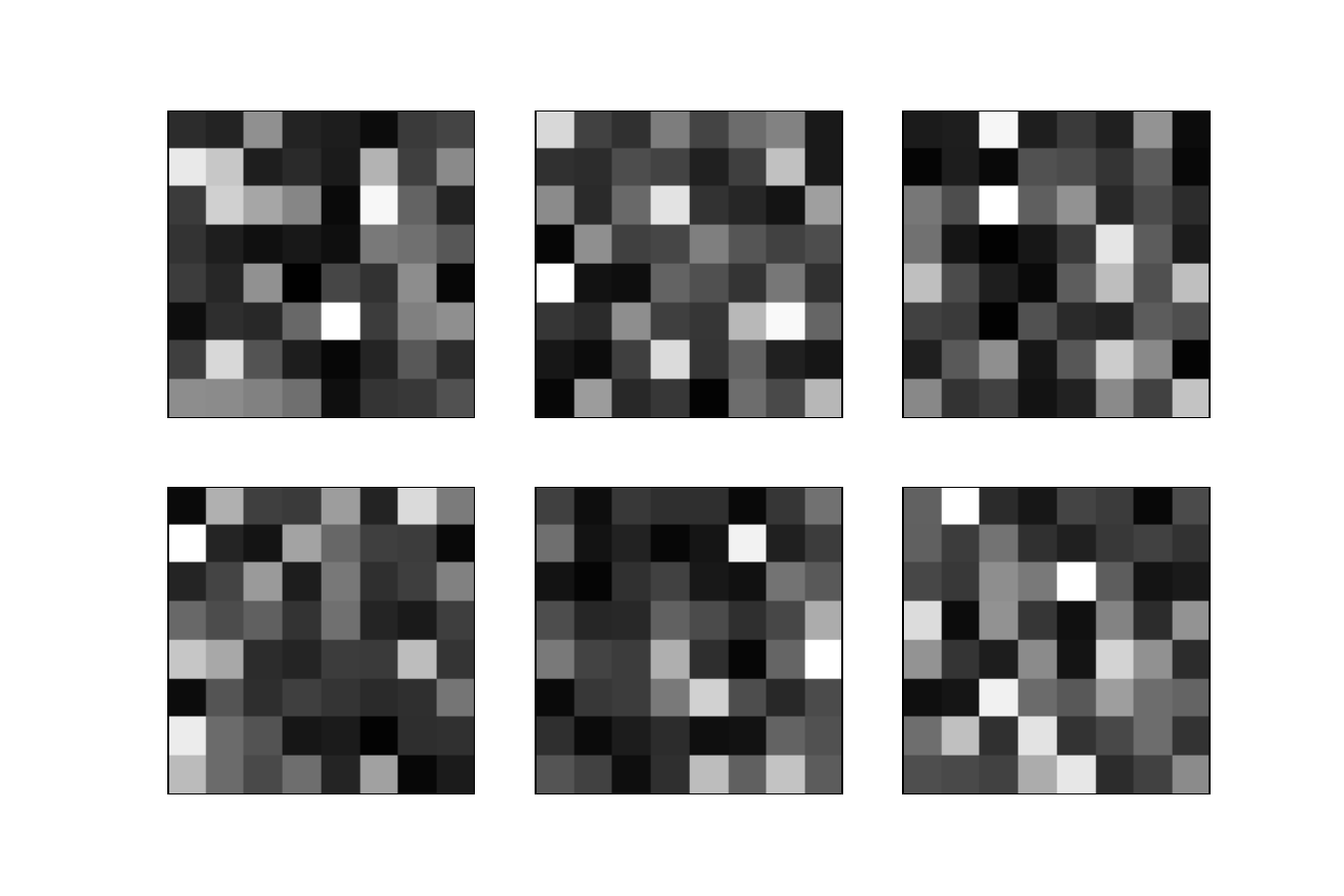} & 
\includegraphics[width=1\linewidth]{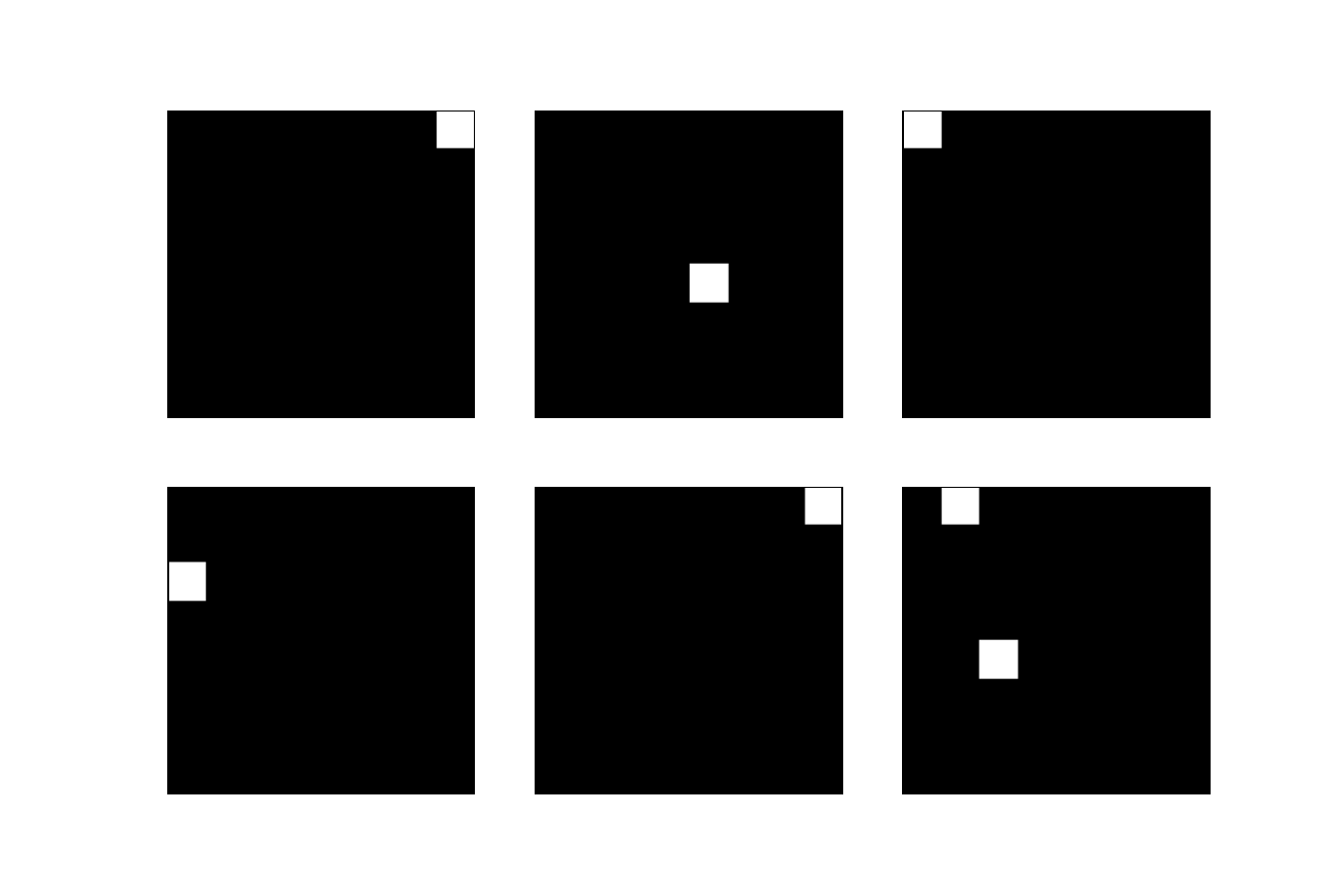} \\
\end{tabular}
\caption{\Xpolish{Optimizations on $\B{M}$ initialized with identity and Hadamard matrices are studied under the AGN and Poisson noise. $\B{M}$ prefers coding strategies under the AGN but favors the raster scan strategy under the Poisson noise regardless of the initialized matrices.}
{
Optimizations on the matrix $\B{M}$, initialized with identity and Hadamard matrices, are explored under both AGN and Poisson noise. While $\B{M}$ demonstrates a preference for strategies with coding under AGN, it consistently leans toward the raster scan strategy under Poisson noise, irrespective of the initialization matrices.
}
}
\label{fig:recon_masks}
\end{figure}
}

\iffalse
\begin{figure}[h]
    \begin{subfigure}{\linewidth}
        \centering
        \caption{ONN optimized masks under Gaussian noise. Left: initialized with the identity matrix. Right: initialized with the Hadamard matrix.}
        % \includesvg[width=1\linewidth]{\homedir/figures/Results/ONN_recon/ONN_recon_gaussian_masks.svg}
        \includegraphics[width=0.49 \linewidth]{\homedir/figures/Results/ONN_recon/ONN_recon_gaussian_masks.pdf}
        \includegraphics[width=0.49 \linewidth]{\homedir/figures/Results/ONN_recon/ONN_recon_gaussian_masks_Hadamard.pdf}
        \label{fig:gaussian_recon_masks}
    \end{subfigure}
    \begin{subfigure}{\linewidth}
        \centering        
        \caption{ONN optimized masks under Poisson noise. Left: initialized with the identity matrix. Right: initialized with the Hadamard matrix.}
        % \includesvg[width=1\linewidth]{\homedir/figures/Results/ONN_recon/ONN_recon_poisson_masks.svg}
        \includegraphics[width=0.49\linewidth]{\homedir/figures/Results/ONN_recon/ONN_recon_poisson_masks.pdf}
        \includegraphics[width=0.49\linewidth]{\homedir/figures/Results/ONN_recon/ONN_recon_poisson_masks_Hadamard.pdf}
        \label{fig:poisson_recon_masks}
    \end{subfigure}
    \caption{First 6 optimized masks}
    \label{fig:recon_masks}
\end{figure}
\fi

\Xhide{
\begin{figure}[h]
    \begin{subfigure}{\linewidth}
        \centering
        \caption{Value distribution of the optimal masks under Gaussian noise}
        % \includesvg[width=1\linewidth]{\homedir/figures/Results/ONN_recon/Mask_value_dist_gaussian.svg}
        \includegraphics[width=1\linewidth]{\homedir/figures/Results/ONN_recon/Mask_value_dist_gaussian.pdf}
        \label{fig:gaussian_recon_masks_dist}
    \end{subfigure}
    \begin{subfigure}{\linewidth}
        \centering
        \caption{Value distribution of the optimal masks under Poisson noise}
        % \includesvg[width=1\linewidth]{\homedir/figures/Results/ONN_recon/Mask_value_dist_poisson.svg}
        \includegraphics[width=1\linewidth]{\homedir/figures/Results/ONN_recon/Mask_value_dist_poisson.pdf}
        \label{fig:poisson_recon_masks_dist}
    \end{subfigure}
    \caption{Masks values distribution (x-axis: mask values; y-axis: counts). (\subref{fig:gaussian_recon_masks_dist}) Given a Raster scan mask family, ONN returns a different mask family under the AGN, indicating that Raster scan is always suboptimal. (\subref{fig:poisson_recon_masks_dist}) ONN adheres to the Raster scan, suggesting it is optimal under Poisson noise.
    textcolor{red}{these can be much smaller}
    }
    \label{fig:masks_value_dist}
\end{figure}
}

\Xhide{\begin{figure}[h]  
\caption*{Distributions of ONN reconstructed masks}
    % \begin{subfigure}{0.49\linewidth}
    % \centering  
    % \caption{$\B{M}_G^I$ VS $\B{M}_P^I$}
    % \includegraphics[width=1\linewidth]{\homedir/figures/Results/ONN_recon/Mask_value_dist.pdf}    
    % \end{subfigure}    
    % \begin{subfigure}{0.49\linewidth}
    % \centering  
    % \caption{$\B{M}_G^H$ VS $\B{M}_P^H$}
    % \includegraphics[width=1\linewidth]{\homedir/figures/Results/ONN_recon/Mask_value_dist_Hadamard.pdf}
    % \end{subfigure}   
    \centering
    \includegraphics[width=.8\linewidth]{\homedir/figures/Results/ONN_recon/Mask_value_dist.pdf}
        \caption{\Xpolish{Denote masks initialized with the identity matrix and optimized under the AGN as $\B{M}_G^I$, masks initialized with the identity matrix and optimized under the Poisson noise as $\B{M}_P^I$, masks initialized with the Hadamard matrix and optimized under the AGN as $\B{M}_G^H$, masks initialized with the Hadamard matrix and optimized under the Poisson noise as $\B{M}_P^H$.  
        $\B{M}_G^I$ is going to increase the light throughput by openning more pixels, but $\B{M}_P^I$ is fixed as it is already an ideal choice. $\B{M}_P^H$ is approaching the identity matrix, differently from the $\B{M}_G^H$. Due to the fact that this optimization problem is not perfectly convex, it is difficult to optimize $\B{M}_P^H$ to the perfect identity matrix, but it shows essential differences between the optimization questions under these two distinct noise models.}{ The Poisson noise data is displayed topside-down by negating its counts. Let $\B{M}_G^I$ represent masks initialized with the identity matrix and optimized under the AGN, and $\B{M}_P^I$ denote masks initialized with the identity matrix and optimized under Poisson noise. Additionally, $\B{M}_G^H$ and $\B{M}_P^H$ represent masks initialized with the Hadamard matrix and optimized under AGN and Poisson noise, respectively. While $\B{M}_G^I$ dynamically enhances light throughput by opening more pixels, $\B{M}_P^I$ remains fixed, representing an already ideal choice. Notably, $\B{M}_P^H$ tends to approach the identity matrix, in contrast to the behavior of $\B{M}_G^H$. The challenge arises from the non-convex nature of this optimization problem, making it difficult for $\B{M}_P^H$ to converge perfectly to the identity matrix. Nevertheless, this deviation underscores fundamental distinctions in optimizing masks under these distinct noise models.
        }
    }
    \label{fig:masks_value_dist}
\end{figure}
}

\Xhide{\bnote{talk about onn first in the last section and then show the reconstruction reuslts before the classifiction}}

\Xhide{
\Xpolish{The different results of the optimal masks suggest the masks optimized under the AGN model may not have the same performance under the Poisson noise. If the measured data is Poisson distributed, then the methods proposed under the AGN assumption are problematic.}
{The results from optimizing the masks indicate that the optimal masks under the AGN model may not perform similarly under Poisson noise. Specifically, methods developed under the AGN assumption may not be effective when the measured data follows a Poisson distribution. \YLnote{For instance, coding optimization based on minimizing the mutual coherence in compressed sensing \cite{mordechay2014matrixRIP} may be defective.} The inherent differences in the underlying noise models of the two distributions can lead to suboptimal performance of AGN-optimized masks in Poisson noise scenarios. Therefore, it is crucial to consider Poisson noise when designing coding schemes and algorithms for imaging applications, especially when working with state-of-the-art sensors.}
}

% move this section to appendix 
%\input{\homedir/sections/4_methods/sensing_matrices_by_prior.tex}

% move this section to appendix 
% \input{\homedir/sections/4_methods/constraint_implement.tex}

\subsection{Computer Simulated Experiments}

\Xpolish{Two classification tasks are introduced in this section. We first test the performances of all the coding schemes on the MNIST dataset for number recognition, and then test them on the hyperspectral data.}{
In this section, we present two distinct classification tasks. 
Initially, we evaluated the effectiveness of various coding schemes using the MNIST dataset, focusing on their performance in number recognition. 
Subsequently, we extend the evaluation to hyperspectral data, examining the adaptability and efficacy of the coding schemes in this different domain.
}

\subsubsection{MNIST Simulated Data Preparation}

% \subsubsection{MNIST Data Pre-processing}

\Xpolish{The MNIST dataset consists of handwritten digits from 0 to 9, with a default size of 28 by 28 pixels \cite{lecun1998MNIST}. In order to align the data with the Hadamard matrices, we padded the images with black pixels at the edges to resize them to 32 by 32 pixels \cite{lecun1998MNIST}. We then transformed the range of all the pixel values from $[0,1]$ to $[0.3, 1]$. This operation enhanced the persuasiveness of the dataset for various noise models, as the added black pixels do not generate Poisson noise, whose variance is proportional to the expected photon counts. \bnote{did we ever vary the dark level (0.3)/ have results showing its effect?}}{
The MNIST dataset, which consists of handwritten digits ranging from 0 to 9, has a default size of 28 by 28 pixels \cite{lecun1998MNIST}. 
To align the data with Hadamard matrices, we add black pixels to the edges of the images and resize them to 32 by 32 pixels \cite{lecun1998MNIST}. 
Subsequently, we rescale the pixel values from the original range of $[0,1]$ to $[0.3, 1]$. This rescaling introduces non-zero Poisson noise to the black regions, which has variance proportional to the expected photon counts.
}

{
We evaluate our proposed model through simulations \Xpolish{that test}{incorporating} both AGN and Poisson noise models, using varying exposure times to generate results at different noise levels. 
We also test all {compressible} strategies, varying the total number of masks but fixing the total exposure time, and define a metric called compression ratio as the ratio of the number of masks to the number of pixels \cite{stojek2022define_compression_ratio}. 
For RS and HB, this ratio is restricted to 1.00, while the rest are evaluated over values in $\mathcal{C}_R \in \{0.01, 0.04, 0.09, 0.16, 0.25, 0.36, 0.49, 0.64, 0.81, 1.00\}$ so that $\B{M}$ of each coding scheme is not restricted to being full-rank. 
The final comparison involves only the best results among all compression ratios for each strategy. Unlike pre-defined masks used by other strategies, the initial masks for ONN can affect performance in the presence of noise. 
We therefore initialized its masks with the PCA components from the training data.
}

\subsubsection{Spectral Datasets and Performance Evaluation}

% \rnote{How well this model can work w/o noise}

% \subsubsection{Performance Evaluation}

\Xhide{\Xpolish{The performances of the models were evaluated through their average classification rates. To ensure a thorough assessment, the model was tested five times independently, each time with a randomly split dataset for training 70\% and validation 30\%. During each assessment, the model generated noisy photon counts, which were then used to train the classifier until the validation rate reached a plateau or declined. The best validation rate achieved during the training process was chosen as the upper bound of representative performance for each assessment. The overall performance of the model was determined by averaging the results of all five assessments.}
{The model performances were assessed based on their average classification rates, which were calculated after five independent tests. In each test, the dataset was randomly divided into 70\% for training and 30\% for validation. The model generated noisy photon counts during each assessment, and the training continued until the validation rate reached a plateau or declined. The highest validation rate obtained during the training process was considered as the representative upper bound for each assessment. The overall performance of the model was determined by averaging the results of all five assessments. This comprehensive evaluation ensured the reliability and accuracy of the model performance.}}

\begin{figure}[ht]
\begin{subfigure}{\linewidth}
    \caption{Indian Pines Visualization}
    \centering
    \includegraphics[width=1\linewidth]{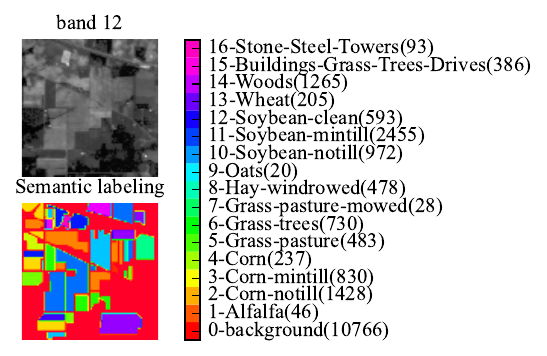}
    
    \label{fig:pines_dataset}
\end{subfigure}
% \end{figure}

% \begin{figure}[h]
\begin{subfigure}{\linewidth}
\caption{Spectrum View of Indian Pines}
    \centering
    % \includegraphics[width=.8\linewidth]{\homedir/figures/Results/Hyperspectra/IndianPines_histogram.png}
    % \caption{Spectrum for each class in the Indian Pines dataset}
    \includegraphics[width=1\linewidth]{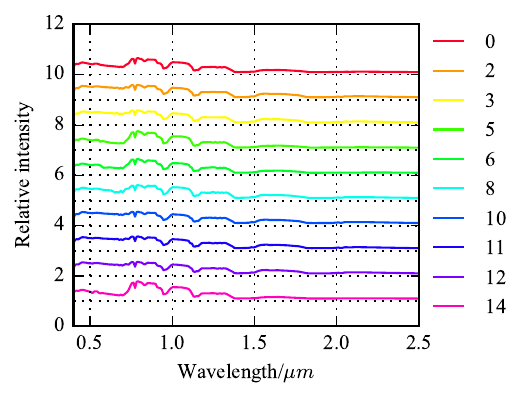}
    % \caption{Average spectrum for each class in the Indian Pines dataset. Classes with data entries fewer than 400 are excluded.}
    \label{fig:AeroRIT_Histogram}
\end{subfigure}
\caption{\Xpolish{Indian Pines dataset visualization\cite{PURR1947}. The top-left figure shows the band 12 of the entire spectrum. The bottom left figure shows the classes by different colors in the legend. The legend includes the class numbers and the corresponding counts.}{(\subref{fig:pines_dataset}) Visualization of the Indian Pines dataset using Band 12 from the entire spectrum (top-left). The bottom-left figure represents classes with distinctive colors as specified in the legend, which includes class labels and their respective counts \cite{PURR1947}.} {(\subref{fig:AeroRIT_Histogram}) Average spectrum for each selected class in the Indian Pines dataset. Classes with data entries fewer than 400 are excluded. Baselines are shown as dashed horizontal lines under each spectrum.}}
\end{figure}

\Xpolish{\Xpolish{In this experiment, w}{W}e utilized the Indian Pines dataset \cite{PURR1947}, which is acquired through the Airborne Visible/Infrared Imaging Spectrometer (AVIRIS) on June 12, 1992, covering the Purdue University Agronomy farm northwest of West Lafayette and its surrounding area. Comprising $145 \times 145$ pixels and 224 spectral reflectance bands in the wavelength range of 0.4 to 2.5 $\mu m$, this dataset provides a comprehensive view of the study area. Fig. \ref{fig:pines_dataset} illustrates the classes within the Indian Pines dataset, presenting the counts associated with each class. The figure also features a representation of one channel (band 12) for all pixels, along with the ground truth of semantic labeling.}{
% text starts
We use the Indian Pines dataset \cite{PURR1947}, acquired through AVIRIS on June 12, 1992, covering the Purdue University Agronomy farm and surrounding areas. 
With $145 \times 145$ pixels and 224 spectral bands (0.4 to 2.5 $\mu m$) per pixel, this dataset offers a comprehensive view of this area. 
Fig. \ref{fig:pines_dataset} illustrates 16 classes in Indian Pines, along with their counts and semantic labeling of the ground truth. 
It features a representation of band-12 for all pixels. 
Fig. \ref{fig:AeroRIT_Histogram} shows the relative intensity of the selected classes from the spectrum data, providing a concise overview of their differences.
}

\Xpolish{In this experiment, spectral data were zero-padded from 224 to 256 to align with the Hadamard mask and \YLnote{\Xpolish{normalized}{re-scaled}} to the range $[0,1]$. We utilized the scanner-classifier module in Fig. \ref{fig:model_config}, adapting the ONN classification layers to the new dataset sizes. Each classification ran for 2000 epochs, with a learning rate of $5 \times 10^{-3}$ and a batch size of 5000.}{
% text
\Xpolish{In the experiment, spectral data \YLnote{extracted as arrays from each pixel}, zero-padded from 224 to 256 for Hadamard mask alignment and rescaled to $[0,1]$, undergoes processing using the modified scanner-classifier module from Fig. \ref{fig:model_config}}{
In the experiment, spectral data vectors, extracted as arrays from each pixel, are zero-padded from 224 to 256 in length for Hadamard mask alignment and rescaled to $[0,1]$. 
It undergoes processing using the modified scanner-classifier module illustrated in Fig. \ref{fig:model_config}}. 
The classifier is \Xpolish{streamlined}{simplified} into a single fully connected layer with 1478 neurons, using the default architecture provided by \texttt{scikit-learn} during the preproject testing phase which gives an 86\% classification rate without any coding or noise. 
The ONN classification layers are adjusted to accommodate input data subsets, each containing 400 entries per class. 
Classes with fewer than 400 entries are omitted to maintain an unbiased input. 
}

\Xpolish{The model performances were assessed based on their average classification rates, which were calculated after five independent tests. In each test, the dataset is randomly split into 90\% for training and 10\% for validation. The model generated noisy photon counts during each assessment, and the training continues until the validation rate reached a plateau or declined. The highest validation rate obtained during the training process is considered as the representative upper bound for each assessment. The overall model performance is determined by averaging the results from all five assessments. This comprehensive evaluation ensures the reliability and accuracy of the model performance.}{
% text
Model performance is evaluated using average classification rates from five independent tests. 
Each test involves a random 90\%-10\% training-validation split of the dataset. Training continues until the validation rate plateaued or declined, with the highest validation rate representing the upper bound for each test. 
The final performance metric is the average of the five maximum validation rates, ensuring a reliable and accurate assessment of the model.
}

\subsection{Hardware Experiments}

\begin{figure}[ht]
\caption*{Hardware Setup}
    \centering    
        \includegraphics[width=.8\linewidth]       {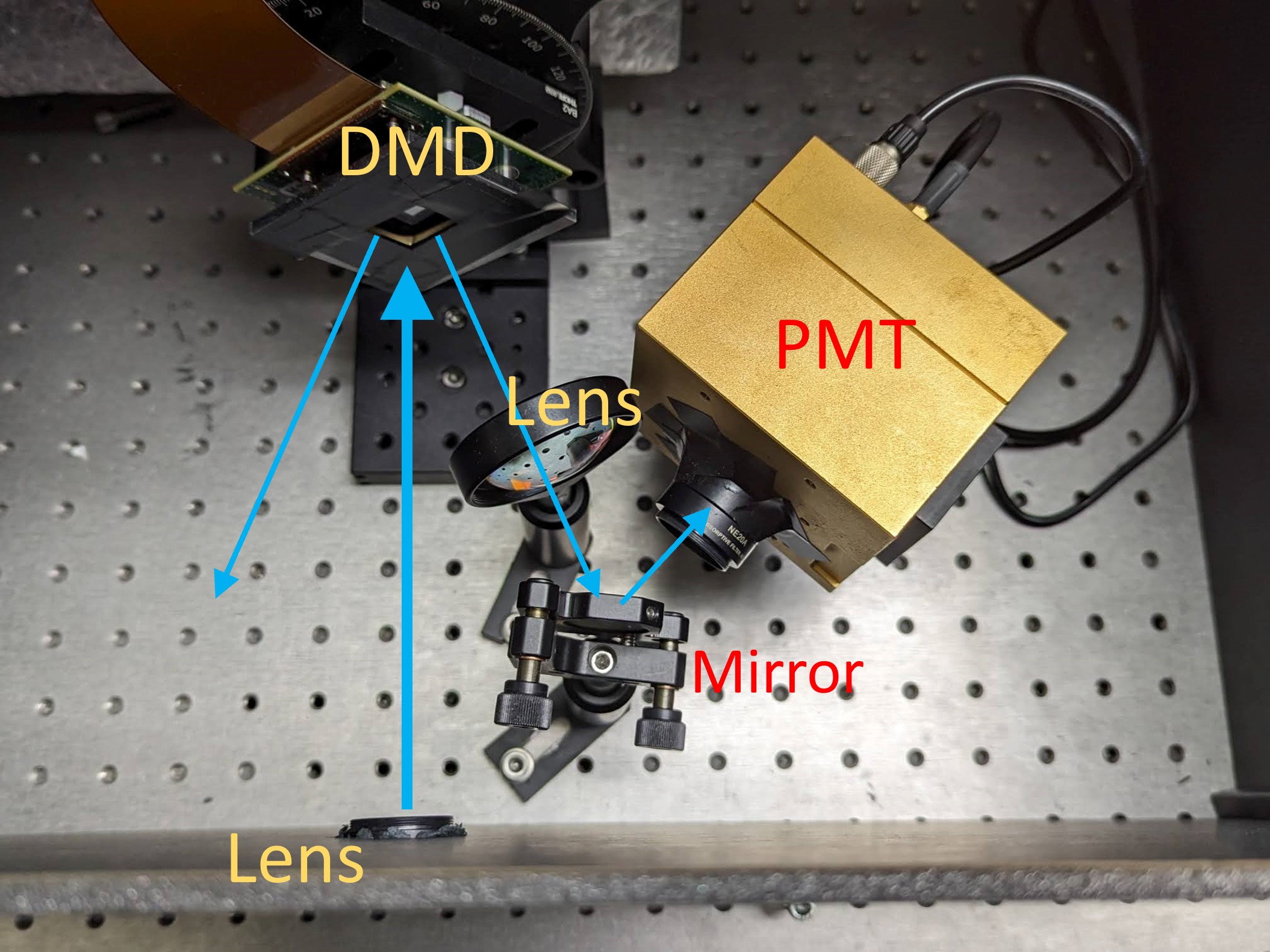}  
        % crop2 is better one
        
    \caption{Experimental configuration of the single-pixel camera: The light path is illustrated by the blue arrows in the diagram. Specifically, our setup utilizes only one branch reflected by the DMD.}
    \label{fig:hardware_setup}
\end{figure}

\Xpolish{Besides the simulations, an experiment is also conducted to enhance our conclusion. The DMD we use is the DLP 7000 purchased from Texas Instruments and the PMT is the PMA Series from PicoQuant. This experiment consisted of a hardware-end data capture and a software-end classifier training. \Xhide{The steps are introduced in the following sections.}}{
% test
To further support our conclusions, we conduct an experiment in addition to simulations. 
As shown in Fig. \ref{fig:hardware_setup}, the experiment utilizes a DMD (digital micromirror device), specifically the DLP 7000 model from Texas Instruments, and a PMT (photomultiplier tube) from PicoQuant, specifically the PMA Series. 
The experimental setup involves both hardware-end data acquisition and software-end classifier training.} 
Handwritten number images are displayed on a monitor, captured by this vision system, and subsequently identified according to the depicted numbers.

% \subsubsection{Data preparation}

\begin{figure}[ht]
    \centering
    % \begin{subfigure}{\linewidth}
    % \centering
    \begin{tabular}{>{\centering\arraybackslash}m{0.3\linewidth}>{\centering\arraybackslash}m{0.2\linewidth}>{\centering\arraybackslash}m{0.3\linewidth}}
         \includegraphics[width=1\linewidth]       {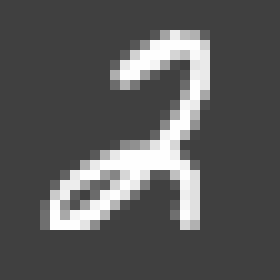}&
        $\B{X} \overset{g(\cdot)}{\longrightarrow} \B{X}^g$&
        \includegraphics[width=1\linewidth]{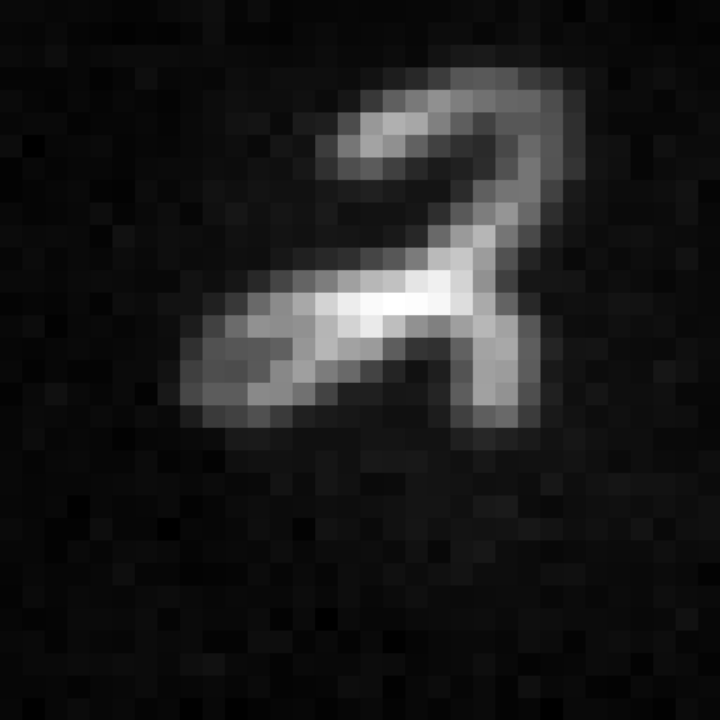}\\
    \end{tabular}
        
        % crop2 is better one        
    % \end{subfigure}
    \caption{Raw image $\B{X}$ (left) VS PMT-observed image $\B{X}^g$}
    \label{fig:sensor_perception}
\end{figure}

\Xpolish{Before data acquisition, we transformed and re-calibrated the MNIST images into what the PMT exactly "observed", and trained the classifier with this new data. This step is important if we don't use the hardware-end data for training.
}{
To ensure that the classifier is properly trained using the PMT data, we perform a transformation $g(\cdot)$ that incorporates \Xpolish{re-calibration}{illumination correction and calibration} of the MNIST images $\B{X}$ to match what the PMT would actually "observe", $\B{X}^g$, during data acquisition. 
Mathematically, this transformation can be described as $\B{X}^g = g(\B{X})$. 
This step aims at reducing the model inconsistency between optimizing the sensing matrices $\B{M}$ and classifiers with digital data on a computer and testing the models with the data acquired from the experiments. 
\Xpolish{This step is crucial when training the classifier without the use of hardware-end data. By implementing this transformation and re-calibration process, we were able to accurately simulate the imaging conditions that the PMT would encounter during actual data acquisition, allowing for more reliable and robust classification results.}{
\Xhide{We choose a subset $\B{X}_s$ from $\B{X}$ and estimated the $g(\cdot)$ between the it and its observed version $\B{X}^g_s$\Xhide{ to ensure that the digital data closely matched the sensor's observations}.}
Specifically, we perform a raster scan of a sample from  $\B{X}_s$, a random subset from $\B{X}$ , with each mask exposed for 1 second. 
The light intensity is adjusted to ensure that the brightest pixel generated approximately 1000 photon counts per second. 
The data $\B{X}_s^g$ captured during this experiment is then reshaped into a 32 by 32 image, as shown in Fig. \ref{fig:sensor_perception}, and $g(\cdot)$ is then \Xpolish{attained}{determined}. 
\Xpolish{After applying this transformation to all digital data \Xpolish{to attain}{to obtain} $\B{X}^g$, \Xpolish{we used it to train the software-end classifier through simulation. }{
we input it into models with all the coding schemes to evaluate their classification performances.}}{}}
{Applying the transformation to digital data yields $\B{X}^g$, which is then fed into models featuring various coding schemes for performance evaluation. }
}

\Xpolish{Finally, we uploaded the masks as the pattern sequences onto the DMD to acquire the data $\noisy{\B{y}}$. The testset consisted of ten images and each handwritten number appeared exactly once. The exposure time for each mask was 1 second. Since all photon records preserved the arrival time, we could change the noise level by choosing random intervals during the whole exposure time and counting the photons within it. }{
\Xpolish{After generating the mask pattern sequences, we upload them onto the DMD to acquire the data, denoted as $\noisy{\B{y}}$.}{
% text starts
Following the generation of mask pattern sequences, they are uploaded onto the DMD to capture data, represented as $\noisy{\B{y}}$.} 
Our test set consists of ten handwritten images, each of which \YLnote{\Xpolish{\AVnote{past tense}appeared}{appears}} exactly once during the experiment. 
To control the noise level in the data, we varied 
% the number of photons recorded within 
lengths of random intervals throughout the 1-second exposure time for each mask. For instance, considering only the photons detected in the first 0.5-second interval, we effectively double the noise level. By preserving the arrival times of all photons, we can systematically adjust the level of noise in the acquired data.
}

\Xhide{
% ignore
\Xpolish{First, we conducted a Raster scan on a sample from MNIST data and each mask in this scan was exposed for 1 second. The light level had been adjusted and the brightest pixel in this experiment generated about 1000 photon records per second. The captured data vector was then reshaped into a 32 by 32 image, which can be treated as a linearly transformed version of the raw image.}
{First, we perform a Raster scan of a sample from the MNIST dataset, with each mask exposed for 1 second. The light intensity was adjusted to ensure that the brightest pixel generated approximately 1000 photon counts per second. The data captured during this experiment was then reshaped into a 32 by 32 image, which can be regarded as a linearly transformed version of the original raw image.}
}

\Xhide{\Xpolish{Then, we estimated the linear transformation between the digital data and exprimental data, and then transformed all digital data in the same way so that the new data are similar to the sensor's observation. This transformed data was then input into the code modules used for simulation to train the software-end classifier. For the selective sensing, the masks were also trained in this step.}
{Then, we estimated the linear transformation between the digital and experimental data to ensure that the digital data closely matched the sensor's observations. After applying this transformation to all digital data, we used it to train the software-end classifier through simulation. Additionally, in this step, the masks used for Selective Sensing were also trained.} \YLnote{The masks with decimals use the fraction of a super pixel \cite{spall22hybrid_training} The ONN training is called \textit{in silico} method (Wright et al) \cite{spall22hybrid_training, wright2021backprop}}}

% \bnote{The ONN should be introduced here first, then show the results in reconstruction}

\section{Results}
% \label{sec:5_results}

%%%%%%%%%%%%%%%%%%%%%%%%%%%%%%%%%%%%%%%%%%%%%
% \input{\homedir/sections/5_results_theory}

% \subsection{Mask optimization using PCA}
% \ynote{Describe
% Show image created wit PCA vs raster for noise levels
% }

% \YLnote{The images of PCA and raster doesn't depend on noise levels}

\subsection{Classification Rates on Simulated Experiment Data}

\Xhide{
\Xpolish{In simulations, both the AGN and Poisson noise models were tested with the model we proposed. The results were generated under different noise levels by changing the exposure time. In addition, all compressible strategies were tested while changing the total number of masks. We defined a metric called compression ratio which is the ratio of the number of masks over the number of pixels. For the RS and HB, this ratio is constrained to be $1.00$. And for the rest, all values in $\mathcal{C}_R = \{0.01, 0.04, 0.09, 0.16, 0.25, 0.36, 0.49, 0.64, 0.81, 1.00\}$ were evaluated. Different from other strategies whose masks were pre-defined, the initial masks for ONN can affect the performances when noise exists. In this project, we initialized its masks by the PCA components from the training data as it usually has the best performances.}
{We evaluated our proposed model through simulations that tested both AGN and Poisson noise models, using varying exposure times to generate results at different noise levels. We also tested all {compressible} strategies, varying the total number of masks, and defined a metric called compression ratio as the ratio of the number of masks to the number of pixels \cite{stojek2022define_compression_ratio}. For RS and HB, this ratio was constrained to be 1.00, while the rest were evaluated over values in $\mathcal{C}_R = \{0.01, 0.04, 0.09, 0.16, 0.25, 0.36, 0.49, 0.64, 0.81, 1.00\}$. Unlike pre-defined masks used by other strategies, the initial masks for ONN can affect performance in the presence of noise. We therefore initialized its masks with the PCA components from the training data, which typically yield better performance.}
}

\begin{figure*}[ht]
    \begin{subfigure}{.49\linewidth}
        \centering
        \caption{Classification under Gaussian Noise}
        % \includesvg[width=1\linewidth]{\homedir/figures/Results/Simulated/gaussian_classification.svg}
        \includegraphics[width=1\linewidth]{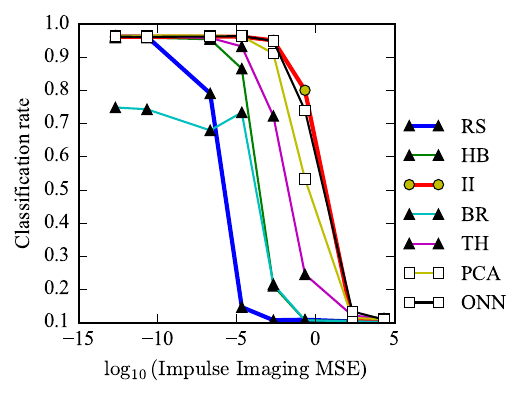}
        \label{fig:gaussian_classification}
    \end{subfigure}
    \begin{subfigure}{.49\linewidth}
        \centering
        \caption{Classification under Poisson Noise}
        % \includesvg[width=1\linewidth]{\homedir/figures/Results/Simulated/poisson_classification.svg}
        \includegraphics[width=1\linewidth]{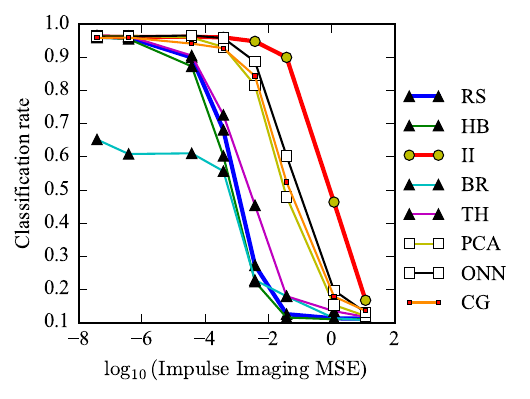}
        \label{fig:poisson_classification}
    \end{subfigure}
    \caption{Classification rates on \textbf{simulated data}. RS: Raster basis. HB: Hadamard basis. II: Impulse imaging. BR: Binary random basis. TH: Low Frequency Truncated Hadamard basis. PCA: PCA basis. ONN: Optical Neural Networks. CG: control group. Selective strategies are marked by $\square$ and non-selective strategies are marked by $\blacktriangle$.(\subref{fig:gaussian_classification}) RS is the worst but using other non-selective coding strategies can improve the performance under the AGN. (\subref{fig:poisson_classification}) Among all non-selective strategies, only selective strategies can significantly improve the performance under Poisson noise. The CG employs the optimized masks under the AGN and trains the software classifier under Poisson noise. Its performance is worse than the ONN.}
    \label{fig:simulated_classification}
\end{figure*}

\Xpolish{The figure \ref{fig:simulated_classification} shows the classification performances with different parameters. The x-axis is the log-scaled mean square error (MSE) of impulse imaging (II) for some exposure time. The measured impulse imaging data requires no reconstruction and can be directly used to compute the noise level via $\|\noisy{\B{y}}_{II} - \B{x}\|_2^2$. As the golden standard in this project, this MSE can work as a metric showing the noise level for both AGN and Poisson noise models. The y-axis is the classification rate on the validation set. The compression ratio can affect the classification performances, so in this figure, we tried all the values in $\mathcal{C}_R$ and picked the best one for TH, PCA and ONN.}{
% text
Fig. \ref{fig:simulated_classification} displays the classification performance of different coding schemes. 
Since Poisson noise is signal dependent and AGN is not, comparing the performance under both models is not straight forward. 
To provide a common metric for both models, whe x-axis is the log-scaled MSE of impulse imaging (II) with varying exposure times. 
\Xhide{Since the measured impulse imaging data requires no reconstruction, the noise level can be directly computed by $\|\noisy{\B{y}}_\mathrm{II} - \B{x}\|_2^2$.} 
This MSE is the gold standard in this project and serves as a metric to measure the noise level for both AGN and Poisson noise models. 
The y-axis represents the validation set's classification rate. 
\Xhide{We explored all the values in $\mathcal{C}_R$ to determine the best compression ratio for TH, PCA, and ONN since compression ratio affects classification performance. Each data point in the Figure represents the best results among all compression ratios for each strategy.}
}
\Xpolish{In this figure, we can see the II always has the best performance, which meets our expectation. When the noise is AGN, using any coding gives rise to a significant improvement compared with RS. For this classification task, using low-rank measurement (HB, PCA, ONN) can further improve this model's classification performances. Notably, the ONN has a performance close enough to the II with only 0.1\% number of photons.}{
% text
A good way to understand the results is to consider the performance of different schemes in relation to the curves for II an RS. 
\YLnote{\Xpolish{Datasets}{Measurements}} for each of the of the schemes can be created as some subset of an II measurement. 
Thus it should not be possible for any coded single pixel system to perform better than II. 
RS is the trivial code allowing simple implementation that also transmits the least amount of light. 
We expect it to perform poorly and any code with worse performance than it seems hard to justify as a useful scheme. 
We evaluate the codes by noting how they perform within these two extremes. 
The results presented in the figure demonstrate that as expected II consistently achieves the highest performance. 
In the presence of AGN, all coding schemes exhibit significant improvements over RS and in many cases provide adequate performance. 
This is because under Gaussian noise, projection from a captured basis into a basis of sparsity where the classifier operates can be done without a significant noise penalty. 
Moreover, for the classification task at hand, \YLnote{\Xpolish{low-rank measurement techniques such as HB, PCA, and ONN}{SS measurement techniques such as PCA and ONN}}, can yield further improvements in classification performance. 
Notably, the ONN approach produces results that are nearly as good as the II approach, \YLnote{\Xpolish{while using only 0.1\% {of the photons} \AVnote{AGN, so no photons}}{while reducing the number of pixels on the sensor from several thousands to one}
}.
}

\Xhide{\bnote{color compressible and non-compressible differently}}

\Xpolish{However, when the measured data is Poisson distributed, HB is no better than RS any more, which agrees the conclusions of Harwit et al. \cite{harwit1979hadamard} that we should not adopt coding for Poisson noise. Though Hadamard coding is not promising, we can still have extra performance gain by using low-rank methods. Among all low-rank methods, the ONN is still the best, suggesting Selective Sensing is the most optimal method for both noise models.}{
However, when dealing with Poisson-distributed measured data, the effectiveness of \Xpolish{Hadamard-based (HB)}{non-selective} coding is no longer superior to that of RS. This aligns with the findings of Harwit et al. \cite{harwit1979hadamard}, which discourage the adoption of coding for Poisson noise. Despite this, there is still potential for performance improvement by utilizing \YLnote{\Xpolish{low-rank}{SS}} methods such as \Xhide{TH, }PCA and ONN. \AVnote{\checkednote{repeating}} \YLnote{\checkednote{This is likely due to the fact that the images with analyze do in fact have sparsity in the Hadamard basis. Selective Sensing where capture patterns are optimized to capture sparsity directly like the ONN can retain their performance under poisson noise making them viable replacements for simple point scanners. This indicates that \Xpolish{Selective Sensing is the most optimal approach for both noise models. }{SS can serve as a potential replacement of II in some scenarios.}}}}

%%%%%%%%%%%%%%%%%%%%%%%%%%%%%%%%%%%%%%%%%%%%%%%%%%%%%%%%%%%
\subsection{Classification Performances on Hardware Experiment Data}

\Xpolish{In the experiments, we used a DMD and a PMT to measure 10 handwritten numbers from the MNIST dataset. The models were trained with $10^5$ photons, which is a greater noise level compared with the experimental environment. The collected data was in the form of a series of timestamps recording the arrival time of the photons. When collecting data, both RS and HB used 1024 masks. The TH and the HB shares the same measured data but the TH only uses the first 92 measured data. PCA and ONN used 92 masks, which is the same as the TH. The physical measurements took 1 second per mask for all strategies. To create data of greater noise level from the raw data, we randomly picked intervals during the 1 second span, and counted the total numbers in it. Since the ONN and PCA only have 92 masks, the maximum exposure time was set as 92 seconds. In a fair comparison, the interval length of each strategy was cpmputed as $\frac{\tau}{m}$ where $\tau$ is the total exposure time we want to investigate and $m$ is number of masks for this strategy.}{
In our experiments, \Xpolish{we utilized a DMD and a PMT to measure ten handwritten digits from the MNIST dataset. To train the models, we employed $10^5$ photons, which constituted a higher noise level in comparison to the experimental setup. T}{t}he collected data take the form of a series of timestamps recording the photon arrival time. 
During data collection, both RS and HB uses 1024 masks, while TH and HB employ the same measured data, but TH uses only the first 92 measurements for compression purposes. Similarly, PCA and ONN utilizes 92 masks. 
Physical measurements take one second per mask for all strategies. To generate data with higher noise levels from the raw data, we randomly select intervals within the 1-second span and count the total number of photons within them. 
Since  ONN and PCA have only 92 masks, the maximum exposure time is set to 92 seconds. 
To ensure a fair comparison, we compute the interval length for each strategy as $\frac{\tau}{m}$, where $\tau$ represents the total exposure time under investigation, and $m$ denotes the number of masks used for each strategy.
}

\Xhide{\bnote{consistent color on figures}}

\begin{figure}[ht]
\caption*{Classification Rates on \textbf{Experimental Data}}
    \centering
    % \includesvg[width=1\linewidth]{\homedir/figures/Results/Experimental_results.svg}
    \includegraphics[width=1\linewidth]{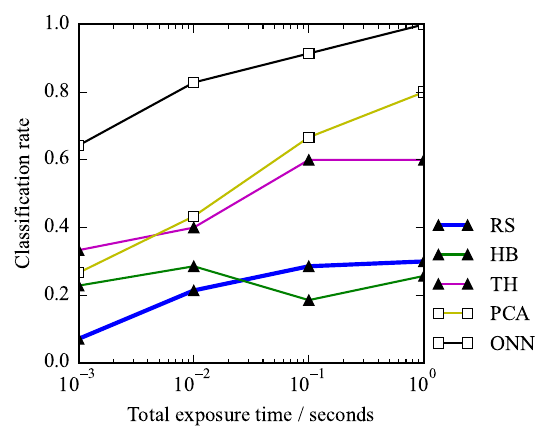}
    \caption{ RS: Raster basis. HB: Hadamard basis. TH: Low Frequency Truncated Hadamard basis. PCA: PCA basis. ONN: Optical Neural Networks. Selective strategies are marked by $\square$ and non-selective strategies are marked by $\blacktriangle$. ONN is in general the best strategies given the experimental data.}
    \label{fig:Exp_classification}
\end{figure}

\Xhide{\bnote{change the x tick labels}}

\Xpolish{However, we still saw some unexpected phenomena in the figure. Firstly, the HB strategy was better than the RS when the exposure was short, which was not observed in the simulation. This phenomenon might arise from the dark photon in the PMT. The dark photon is a signal-independent noise which has about 40-80 counts per second and could usually be observed before the experiments started. When the exposure was short, this noise would be more significant and gave rise to a better performance of the HB. Secondly, the TH achieved better classification performances than PCA when the exposure time is short. It was the consequence of using unneeded PCA masks. To verify, we fixed the toal number of photons at $10^5$, change the compression ratio, and made figure \ref{fig:classification_ratio} by simulation. This figure shows that PCA method can be worse than TH when it has more masks than needed. In other words, PCA strategy requires an careful estimation of the noise level to avoid using extra masks. Notably, the ONN looks more stable when includes extra masks. That can be seen as another advantage of the Selective Sensing. After the models are trained, they may be applied in unknown noise levels. Different from PCA strategy whose optimal number of masks is noise level sensitive, the ONN can have a better and robust performance even when the number of masks is not optimal.}
 {Fig. \ref{fig:Exp_classification} displays the classification rates achieved on the experimental data. The x-axis shows the exposure time, and the y-axis represents the classification rates of the models. Although there may be discrepancies between software training and hardware data acquisition, the ONN models \Xpolish{performed excellently}{outperform other coding schemes across all noise levels}. However, the figure reveals some unexpected phenomena. Firstly, the HB strategy outperforms the RS for short exposure times, which is not observed in the simulations. This phenomenon may have arisen due to the dark counts in the PMT, a signal-independent noise that has about 40-80 counts per second \YLnote{\Xpolish{, and is more pronounced in short exposures}{under our lab condition}}. Secondly, TH achieves better classification performances than PCA when the exposure time is short. This result is due to the use of unneeded PCA masks since the compression ratio was not optimized as it was in the simulation. To verify, we conduct simulations and obtain Fig. \ref{fig:2d_classification_plots} by fixing the total number of photons at $10^5$ and varying the compression ratio under Poisson noise. This figure shows that PCA method can perform worse than the TH if \YLnote{\Xpolish{extra masks are used}{an improper compression ratio is chosen}}. Hence, PCA strategy requires careful estimation of the noise level to avoid \YLnote{\Xpolish{using extra masks}{improper compression ratios}}. \YLnote{\Xpolish{Notably, the ONN appears more stable when including extra masks, which is another advantage of the ONN. The ONN can perform robustly even with suboptimal numbers of masks, unlike PCA strategy, which is noise-level sensitive. While it is possible to set the noise level during training, the effectiveness of the ONN can be evaluated for other levels of noise as well.}{Notably, ONN demonstrates enhanced stability and robustness when confronted with this challenge.}} This flexibility in application allows for broader use of the ONN beyond its specific training conditions, increasing its potential impact and utility in real-world scenarios.}

\Xhide{
\begin{figure}[ht]
    \centering
    \includesvg[width=1\linewidth]{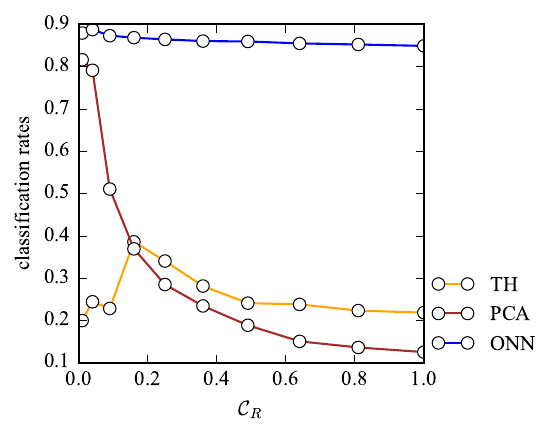}
    \caption{Classification rates VS compression ratio under Poisson noise. TH: Truncated Hadamard basis. PCA: PCA basis. ONN: Optical Neural Networks. Given a certain noise level, PCA is required to be accurate in the mask numbers, or it can be worse than TH. ONN doesn't have such requirement.}
    \label{fig:classification_ratio}
\end{figure}
}

\Xhide{\YLnote{2D plot. \& Don't re-train the model. For training, choose a fixed noise (1e5) level, and for testing, use multiple noise levels}.}

\Xhide{
\begin{figure}[ht]
    \centering
    \includesvg[width=1\linewidth]{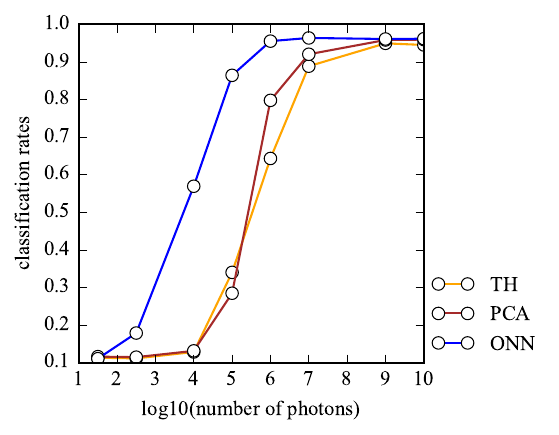}
    \caption{Classification rates VS light level under Poisson noise. TH: Truncated Hadamard basis. PCA: PCA basis. ONN: Optical Neural Networks. Given a certain noise level, PCA is required to be accurate in the mask numbers or it can be worse than TH. ONN doesn't have such requirement.}
    \label{fig:classification_light_level}
\end{figure}
}

\begin{figure*}[ht]
    \begin{subfigure}{0.32\linewidth}
        \centering
        \caption{Truncated Hadamard Classification Rates}
        % \includesvg[width=1\linewidth]{\homedir/figures/Results/2d_contours/low_freq_truncated_hadamard_2d_classification.svg}
        \includegraphics[width=\linewidth]{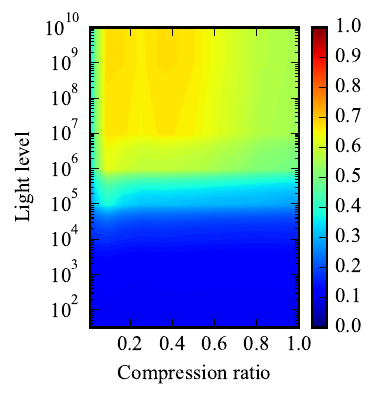}
        \label{fig:truncated_hadamard_2d_classification}
    \end{subfigure}
    \begin{subfigure}{0.32\linewidth}
        \centering
        \caption{Hardware PCA Classification Rates}
        % \includesvg[width=1\linewidth]{\homedir/figures/Results/2d_contours/hardware_pca_2d_classification.svg}
        \includegraphics[width=\linewidth]{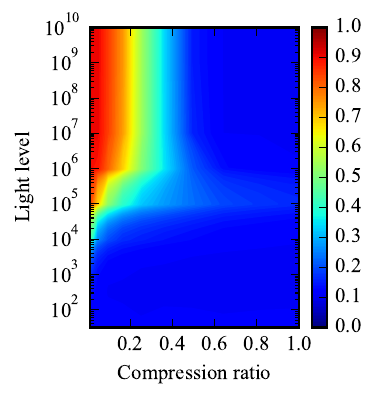}
        \label{fig:hardware_pca_2d_classification}
    \end{subfigure}
    \begin{subfigure}{0.32\linewidth}
        \centering
        \caption{ONN Classification Rates}
        % \includesvg[width=1\linewidth]{\homedir/figures/Results/2d_contours/onn_2d_classification.svg}
        \includegraphics[width=\linewidth]{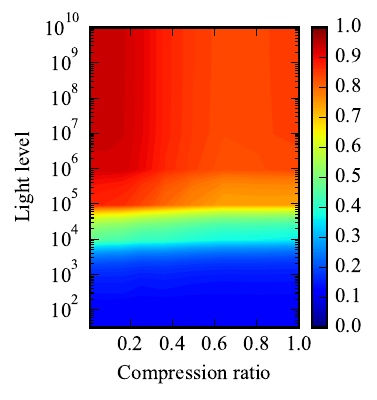}
        \label{fig:onn_2d_classification}
    \end{subfigure}
    \caption{Classification rates on simulated data regarding compression ratios and light levels.  The models are trained under Poisson noise using a light level of $10^5$ photons and subsequently tested under different light levels. It is observed that hardware PCA performance deteriorates when the compression ratio is not properly chosen, whereas the Truncated Hadamard and ONN models exhibit greater robustness in such scenarios.}
    \label{fig:2d_classification_plots}
\end{figure*}

\Xhide{\YLnote{Show: fixed CR and change the noise level}}

\Xhide{\bnote{use $C_R$ on x axis}}

\Xhide{\YLnote{Add how to adjust the experimental data to be classified by simulated model}}

%%%%%%%%%%%%%%%%%%%%%%%%%%%%%%%%%%%%%%%%%%%%%%%%%%%%%%%%%%%
\subsection{Simulated Experiment on Hyper-Spectral Data}

\Xhide{
\Xpolish{For this experiment, we use the Indian Pines dataset \cite{PURR1947}, which was acquired by employing Airborne Visible / Infrared Imaging Spectrometer (AVIRIS) on June 12, 1992 over the Purdue University Agronomy farm northwest of West Lafayette and the surrounding area.
This dataset is composed of $145 \times 145$ pixels and 224 spectral reflectance bands in the wavelength range 0.4 to 2.5  $\mu m$
In Fig. \ref{fig:pines_dataset}, the classes of the Indian Pines dataset are depicted, along with the respective counts within the dataset. The figure showcases an illustration of one channel (band 12) for all pixels, along with the ground truth of semantic labeling.}
{
In this experiment, we utilized the Indian Pines dataset \cite{PURR1947}, which was acquired through the Airborne Visible/Infrared Imaging Spectrometer (AVIRIS) on June 12, 1992, covering the Purdue University Agronomy farm northwest of West Lafayette and its surrounding area. Comprising $145 \times 145$ pixels and 224 spectral reflectance bands in the wavelength range of 0.4 to 2.5 $\mu m$, this dataset provides a comprehensive view of the study area. Fig. \ref{fig:pines_dataset} illustrates the classes within the Indian Pines dataset, presenting the counts associated with each class. The figure further features a representation of one channel (band 12) for all pixels, along with the ground truth of semantic labeling.
}
}

\begin{figure*}[thp]
    \begin{subfigure}{0.5\linewidth}
        \centering
        \caption{Classification under Gaussian Noise}
        % \includesvg[width=1\linewidth]{\homedir/figures/Results/Hyperspectra/classification_vs_light_level_gaussian.svg}
        % \includegraphics[width=1\linewidth]{\homedir/figures/Results/Hyperspectra/classification_vs_light_level_gaussian.pdf}
         \includegraphics[width=1\linewidth]{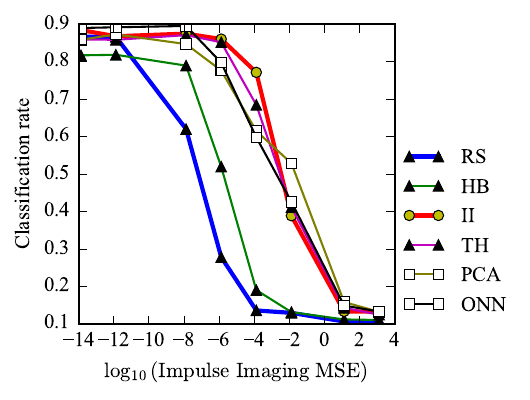}
        \label{fig:pines_gaussian_classification}
    \end{subfigure}
    \begin{subfigure}{0.5\linewidth}
        \centering
        \caption{Classification under Poisson Noise}
        % \includesvg[width=1\linewidth]{\homedir/figures/Results/Hyperspectra/classification_vs_light_level_poisson.svg}
        % \includegraphics[width=1\linewidth]{\homedir/figures/Results/Hyperspectra/classification_vs_light_level_poisson.pdf}
        \includegraphics[width=1\linewidth]{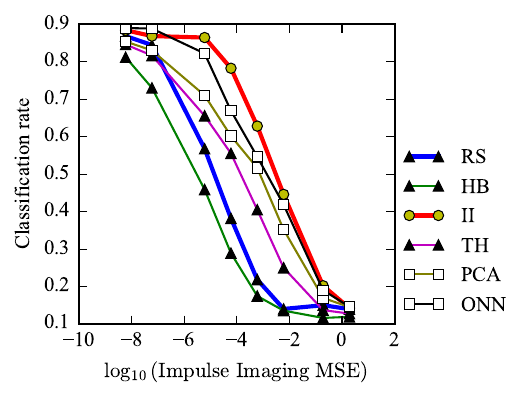}
        \label{fig:pines_poisson_classification}
    \end{subfigure}
    \caption{Classification rates of Indian Pines dataset varying light level. RS: Raster basis. HB: Hadamard basis. II: Impulse imaging. TH: Low Frequency Truncated Hadamard basis. PCA: PCA basis. ONN: Optical Neural Networks.}
    \label{fig:pines_result}
\end{figure*}

\Xpolish{In this experiment, we zero-padded each spectrum from 224 to 256 to align with the Hadamard mask. Spectral data are normalized within the range of $[0,1]$. The scanner-classifier module from Fig. \ref{fig:model_config} was utilized, with modifications made to the ONN classification layers to match the sizes of the new dataset. Each classification was performed for 2000 epochs, with a learning rate of $5 \times 10^{-3}$ and batch size of 5000.
Figure \ref{fig:pines_result} shows the classification rates for the Indian Pines dataset varying light level from $10^{1}$ to $10^{10}$.
% $10^-1$ to $10e^10$.
% To ensure accuracy, each classification was repeated 5 times, and the average was computed as the final result. 
}
\Xpolish{
\Xhide{In this experiment, spectral data were zero-padded from 224 to 256 to align with the Hadamard mask and \YLnote{\Xpolish{normalized}{re-scaled}} to the range $[0,1]$. We utilized the scanner-classifier module from Fig. \ref{fig:model_config}, adapting the ONN classification layers to the new dataset sizes. Each classification ran for 2000 epochs, with a learning rate of $5 \times 10^{-3}$ and a batch size of 5000.} Figure \ref{fig:pines_result} illustrates the classification rates for the Indian Pines dataset across varying light levels\Xhide{ from $10^1$ to $10^{10}$}. Similar to the results on the MNIST dataset, any coding schemes can improve the classification performance when the noise is Gaussianian but only SS can perform as comparably well as impulse imaging when the observed data is Poissonian. This result suggests that SS has the potential to be promoted into hyperspectrum data.
}{
Fig. \ref{fig:pines_result} depicts the classification rates for the Indian Pines dataset under different light levels. 
Consistent with the findings on the MNIST dataset, all coding schemes enhance classification performance under Gaussian noise. 
However, under Poisson noise, only SS demonstrates comparable efficacy to impulse imaging. 
This outcome implies that SS holds promise for advancement in hyperspectral data applications.
}

\begin{figure}[htp]   
% reference cannot display as expected if ended with figure*
\end{figure}

\section{Discussion and Limitations}
\label{sec:6_discussion}

\Xpolish{This paper has shown that the challenges for computational imaging under Poisson noise. Algorithms based on the AGN noise assumption are problematic on the modern sensors. However, if the task requires no reconstruction but direct feature extraction, the Selective Sensing using the Optical Neural Networks model can find the optimal coding methods. We have shown the feasibility of the Selective Sensing via simulations and experiments, and it demonstrated a promising classification performance on the MNIST handwritten number dataset. Also, it is robust in the application scenarios where their noise level is hard to estimate. Furthermore, the Selective Sensing provides an motivation for proposed optical ANNs or ANNs with optical layers which allow us to optimize the coding schemes wherever the Poisson noise happens. }{
Our paper highlights the challenges of computational imaging under Poisson noise and its impact on algorithms based on the AGN noise assumption. 
We find that for compressible measurements, and especially tasks that involve direct feature extraction instead of signal reconstruction, a Selective Sensing approach using task-based optimized codes provides a viable coding solution.
Through simulations and experiments, we demonstrate the feasibility of Selective Sensing and its promising classification performance on the MNIST handwritten number dataset. 
It is also robust in application scenarios with difficult-to-estimate noise levels. 
Our ONN method represents a method that can generate these selective measurements.
Furthermore, Selective Sensing motivates the development of optical ANNs or ANNs with optical layers to globally optimize imaging systems.
}

% \subsection*{\checkednote{Definition of a measurement}}
\iffalse
In \Xpolish{compressed sensing and }{computational imaging}, coding is typically defined as a measurement of an analog flux through some type of coding projection or in our example a mask. As has been shown by multiple works, this analog model of light does not account for the poisson noise inherent in any real measurements and leads to counter intuitive behavior of coding approaches.

If we instead think of our system as measuring photons through different codes, the code behavior makes intuitive sense. A photon measured through a mask with many open pixels carries less information about the scene than one captured through a raster mask because our measurement is ambiguous regarding which pixel in the mask was the origin of the photon. In a raster mask every photon can be uniquely assigned to one pixel. In essence, more photons do not equal more information.

The implication of this well documented problem become ever more important in the age of low noise and photon counting cameras where Poisson noise dominates all measurements. It is wide reaching since the projection process we study here in a specific coding experiment is part of the design of any camera. In other words: Any camera or vision system has to project data from a high dimensional scene space down into a lower dimensional sensor space where it encounters Poisson noise and then uses those noisy measurements to make inferences about the scene. 
\fi 
% replaced by the below
In computational imaging, coding typically involves measuring analog flux through a coding projection such as mask in our case. However, this analog model neglects inherent Poisson noise in real measurements, leading to counterintuitive coding behaviors. Viewing the system as measuring photons through different codes provides more intuitive insights. Both total number of photons and the amount of information about their original pixels matter. For instance, a photon measured through a mask with open pixels carries less scene information than one through a raster mask, where each photon can be uniquely assigned to a pixel. This challenge becomes critical in the era of low-noise, photon-counting cameras dominated by Poisson noise. This issue is significant as the projection process studied in specific coding experiments is integral to the design of any camera, where high-dimensional scene data is projected into a lower-dimensional sensor space encountering Poisson noise for subsequent scene inferences.

\Xpolish{On the other hand, there are some limitations in our project. First, we employed the AGN model and reparameterization trick for the model training, which is only an approximation to the noise at the sensor. Second, our test set in the experiments only contains 10 numbers, which may not be representative enough. There is also an inconsistency as the model was trained by simulated data but tested with experimental data. Last but not the least, the Photon Distribution Factor rescales the masks $\B{M}$, but its value changes during the training and it is not evolved in the back-propagation. In general, the optimization of the ONN model still has some defects and we still need to improve the results by using better optimization methods and more experimental data.}{
% text
Despite the promising results of our project, there are some limitations that must be acknowledged. 
% First, we used a Gaussian noise model with reparameterization to train our model, which is only an approximation of the actual \Xpolish{quantization}{photon} noise at the sensor. 
First, we used a noise model where the variance equals the mean and applied reparameterization to approximate the actual photon noise at the sensor. This approach offers a valid alternative to Poisson noise, which is also just an approximation.
Second, our test set consisted of only 10 numbers, which may not provide a comprehensive evaluation of the model's performance. 
\Xhide{Additionally, we noted an inconsistency in that the model was trained using simulated data but tested with experimental data. 
\Xpolish{Lastly, the \YLnote{Photon Distribution Factor \#no referrence before\#} rescales the masks $\B{M}$}{Lastly, we employ a weighted re-distribution method to satisfy the constraint \ref{constraint:photonNumber} for sensing matrices}, but its value changes during training and is not evolved during back-propagation.}
These limitations highlight the need for further improvements in the optimization of the ONN model, such as using more advanced optimization methods and larger sets of experimental data.
Our work highlights the importance of the integration of imaging hardware and signal processing. 
In single photon accurate imaging systems, comprehensibility and sparsity of the data can be exploited to far greater effect during the measurement, as opposed to post processing.
}

% Any acknowledgments to only be included in camera ready
% \ifpeerreview \else
% \section*{Acknowledgments}
% \fi

\clearpage

\bibliographystyle{IEEEtran}
\bibliography{\homedir/Main/bibtex.bib}
% \bibliography{hardware_pca_onn/Main/bibtex}

\ifpeerreview \else
%%%% For the camera ready version, please fill out this
%%%% biography. Your camera ready should be within a 12 page limit
%%%% including acknowledgments, references and biography.

% \includegraphics command within an optional argument. (You could
% create your own custom macro containing the \includegraphics command
% to make things simpler here.)
% \begin{IEEEbiography}[{\includegraphics[width=1in,height=1.25in,clip,keepaspectratio]{mshell}}]{Michael Shell}

%% BIOGRAPHY
% \input{\homedir/sections/biography}

%\newpage


%\vfill

\fi


\ifCLASSOPTIONcompsoc
  \section*{Acknowledgments}
\else
  \section*{Acknowledgment}
\fi

 The authors would like to thank Dr. Rebecca Willet for her invaluable assistance in refining the problem statement and providing support for our conclusions during the course of this research. Additionally, the authors would like to express gratitude to Dr. Sebastian Bauer for his guidance in understanding the mathematical aspects of this problem at its early stages. This work is supported by the National Science Foundation (1846884) and the Air Force Office for Scientific Research (FA9550-21-1-0341).

 % This work was supported by funding from the National Science Foundation Career grant.



%\bibliographystyle{IEEEtran}


% \end{thebibliography}

\end{document}